\documentclass[a4paper,11pt]{article}
\pdfoutput=1 
\usepackage{jcappub}

\usepackage[T1]{fontenc} 
\usepackage{ae,aecompl}
\usepackage{graphicx}
\usepackage{float}
\usepackage{amsmath}
\usepackage{amssymb}
\usepackage{xcolor}
\usepackage{xspace}
\usepackage{txfonts}
\usepackage{ulem}
\usepackage{hyperref}

\newcommand{\kB}{{k_{\rm B}}}
\newcommand{\me}{{m_{\rm e}}}
\newcommand{\sigT}{{\sigma_{\rm T}}}
\newcommand{\Te}{{T_{\rm e}}}
\newcommand{\Tcmb}{{T_{\rm CMB}}}

\newcommand{\muc}{\mu_{\rm c}}

\newcommand{\betac}{\beta_{\rm c}}
\newcommand{\gammac}{\gamma_{\rm c}}

\newcommand{\thetae}{\theta_{\rm e}}
\newcommand{\Ne}{N_{\rm e}}

\newcommand{\Kz}{K_0}
\newcommand{\Kp}{K'}
\newcommand{\Pz}{P_0}
\newcommand{\Pp}{P'}
\newcommand{\kzbh}{\pmb{\hat{k}}_0}
\newcommand{\kpbh}{\pmb{\hat{k}}'}
\newcommand{\kzb}{\pmb{k}_0}
\newcommand{\kpb}{\pmb{k}'}
\newcommand{\pzb}{\pmb{p}_0}
\newcommand{\ppb}{\pmb{p}'}
\newcommand{\pzbh}{\hat{\pmb{p}}_0}
\newcommand{\ppbh}{\hat{\pmb{p}}'}
\newcommand{\omegaz}{\omega_0}
\newcommand{\omegap}{\omega'}

\newcommand{\gammaz}{{\gamma}_0}
\newcommand{\gammap}{{\gamma}'}
\newcommand{\pz}{p_0}
\newcommand{\pc}{p_{\rm c}}

\newcommand{\betaz}{{\beta}_0}

\newcommand{\muz}{\mu_0}
\newcommand{\mup}{\mu'}
\newcommand{\musc}{\mu_{\rm sc}}

\newcommand{\expf}[1]{{\rm e}^{#1}}

\newcommand{\SZpack}{{\tt SZpack}\xspace}

\newcommand{\D}[1]{\ensuremath{\mathrm{d}{#1}\,}}

\newcommand{\qqquad}{\qquad\qquad}

\title{The SZ effect with anisotropic distributions and high energy electrons}

\author[a]{Elizabeth Lee}
\author[a]{and Jens Chluba}

\affiliation[a]{Jodrell Bank Centre for Astrophysics, School of Physics and Astronomy, The University of Manchester, Oxford Road, Manchester, M13 9PL, U.K.}

\emailAdd{Jens.Chluba@Manchester.ac.uk}

\date{March 2024}

\begin{document}

\abstract{
Future observations of the Sunyaev-Zeldovich (SZ) effect promise ever improving measurements in terms of both sensitivity and angular resolution. As such, it is increasingly relevant to model `higher-order' contributions to the SZ effect.
This work examines the effects of high-energy non-thermal electron distributions and those of anisotropic electron and photon distributions on the SZ signals. 
Analytic forms of the anisotropic scattering kernels for photons and electrons have been derived and investigated. We present a method for determining the anisotropic contributions through a spherical harmonic decomposition to arbitrary angular multipoles, and discuss the behaviour of these scattering kernels. 
We then carry out an exploration of various simplistic models of  high energy non-thermal electron distributions, and examine their anisotropic behaviour.
The kinematic SZ in the relativistic regime is studied using the kernel formulation allowing us to clarifying the role of kinematic corrections to the scattering optical depth.
We finally present a release of an updated and refined version of \SZpack including a new integrated {\tt Python} interface and new modules for the calculation of various SZ signals, including those described in this paper.
}
%--------------------------------------------------------------------------------

\maketitle

%--------------------------------------------------------------------------------
\section{Introduction}
%--------------------------------------------------------------------------------
The Sunyaev-Zeldovich (SZ) effect \citep{Zeldovich1969,Sunyaev1970} is a probe of galaxy clusters and the large-scale structure in our Universe. It emerges as a consequence of hot plasmas upscattering the comparatively cold photons of the cosmic microwave background (CMB) in the Doppler-dominated regime of Compton scattering.\footnote{This means that Klein-Nishina corrections can be neglected throughout.} As such, it leads to unique distortions in the CMB that can be used to increase our understanding of the galaxy clusters alongside the CMB itself.

In this paper, we focus on the effects of anisotropy in the incoming electron and photon distributions alongside an exploration of the effects of simple models of non-thermal high energy electron populations and their intersection with anisotropy. We approach the scattering problem using redistribution kernels that can be applied up to high electron momenta in the relativistic regime.

Overviews of the traditional SZ effect can be found in e.g., \cite{Carlstrom2002} and \cite{Mroczkowski2019}. Generally speaking, the classical thermal SZ (tSZ) effect is derived by assuming an isotropic thermal photon distribution and an isotropic thermal electron momentum distribution. The kinematic SZ (kSZ) effect -- the signal caused by bulk motion of the cluster in the CMB frame -- can be then understood as the two populations being isotropic in their respective frames, while leading to anisotropy in the interaction due to their relative motion. The related scattering problem can then be solved by either using an anisotropic electron distribution to scatter the isotropic CMB \citep{Sazonov1998, Nozawa1998SZ, Challinor1999, Chluba2012} or to apply anisotropic photon scattering kernels in the rest frame of the electron cloud to obtain the results by successive Lorentz transformations \citep{Chluba20122x2, Chluba2012, Nozawa2013}. The latter approach allowed clarifying the role of kinematic corrections to the thermal SZ effect \citep{Chluba2012}; however, only up to second order in the peculiar motion of the cluster. 
The relativistic correction due to the motion of the observer can be added similarly \citep{Chluba2005, Nozawa2005, Chluba2012}.
Here we provide the formalism to extend to arbitrary electron momenta with the  limitation that in the restframe of the scattering electron Klein-Nishina corrections need to remain small.

While the impact of `anisotropies' has been discussed indirectly for the SZ effect, their role is more explicit in the context of the polarized SZ (pSZ) effect \citep[e.g.,][]{Sazonov1999, Hu2004pSZ, Khabibullin2018}.
Importantly, the kSZ effect relies predominantly on induced dipoles in the distributions, while the pSZ relies on the quadrupole \citep{Sunyaev1980, Itoh2000pSZ}. 
%
%The pSZ effect also only evaluates the polarised signal, so discussions generally do not discuss the effect of quadrupole anisotropies on the overall intensity.
%
However, beyond motion-induced anisotropies, the SZ effect may also be caused by locally anisotropic electron distributions (particularly for jets and outflows) or the intrinsic anisotropies in the photon background itself. In these cases, complex scattering occurs between different anisotropies -- as the scattering generally tends towards creating isotropy within the medium.

Here, we discuss a framework to consider arbitrary multipole components to SZ scattering. This follows on the works of e.g., \cite{Ensslin2000, Chluba2014b} to create scattering kernels that describe the probability of a scattering from one frequency to another given a particular incoming electron momentum. We extend the previous considerations to an anisotropic formulation allowing for the expression of either the electron {\it or} the photon distribution in terms of an spherical harmonic decomposition about the line-of-sight. This allows for an accurate computation of the impact of anisotropies in the SZ effect. While the expressions given here should be applicable with high accuracy to the SZ effect, for general problems there may be some aspects still to be uncovered, as we remain in the Doppler-dominated regime instead of extending to general Compton scattering \citep[e.g.,][]{Jones1968, Nagirner1994, Belmont2009, Sarkar2019}. A generalization is left to future studies, but may have a range of applications (see Sect.~\ref{sec:beyond}).

It has been well established that the photon anisotropies generally lead to small signals \citep[e.g.,][]{Sazonov1999, Chluba2014a, Chluba2014b}; however, in the future high precision measurements are expected, such that new effects may come into observational reach. We also highlight that the role of electron anisotropy has been studied far less, as it, for instance, relies on a firm understanding of the magnetic fields within the ICM and other sources of microphysical anisotropies \citep{Hu2004pSZ, Khabibullin2018}. However, these previous studies indicate that this could lead to pSZ effects of a comparable size to the other commonly discussed pSZ effects. It is also worth noting that microphysical electron anisotropies may play a larger role in the non-thermal regime, which is not well-thermalised and may magnify any effects that are generated. This calls for generalized treatments of the scattering problem as we present here with a focus on intensity.

Accordingly, alongside a consideration of anisotropies, we also examine some simple models of high-energy non-thermal electron populations \citep[e.g.,][]{Ensslin2000, Colafrancesco2002,Kaastra2009} to showcase how the related SZ signal differ from those due to thermal electron distributions. It is worth making an aside to clarify what is meant by {\it non-thermal electron populations}. In general, clusters are largely thermalised -- they sit in massive gravitationally-bound halos that are the primary cause of the high cluster temperatures. However, firstly, there is a temperature distribution across clusters, with them (generally-speaking) being hotter closer to the core and cooler towards the edges \citep[see][for discussion in the context of SZ]{Lee2020, Lee2022sims}. Secondly, local non-thermal electron disruption can be caused by a number of processes -- e.g., jets, feedback, shocks, turbulence and cosmic rays, which need to be considered separately.

In general, when there is temperature variance within clusters, it must be determined whether the electron distribution is always locally thermalised, in which case the total SZ effect can be calculated by superposing the contribution from each component along the line of sight within clusters. To simplify the computation, a moment expansion around the mean temperature can be applied, leading to a clean separation of spectral and spatial effects \citep[e.g.,][]{Chluba2013, Lee2020}. On the other hand, when the departures lead to local non-thermal electron distributions, the inherent local SZ distortion will necessarily be modified, and more general methods are required to compute the signal.

A further distinction must also be made about the source of non-thermality. While, for instance, turbulence and `non-thermal' pressure exist in clusters leading to low-energy modifications to the SZ distortion -- the effects on the observed cluster tSZ effect and relativistic corrections will be small (although the effects may be relevant in consideration of the resolved kSZ effect). However, in this paper, we focus on high-energy non-thermal contributions sourced by, for instance, jets and shocks. These could have significant effects on the observed signal, especially in the high-energy tail of the SZ signal. These same high-energy phenomena may also cause over-heated regions outside of the extent of the clusters themselves, that could be observed with high-precision, well-resolved CMB measurements \citep{Malu2017,Acharya2021}, requiring robust treatments of the underlying scattering processes given here.

Finally, this paper also serves to announce the release of an updated and improved version of \SZpack\footnote{\SZpack can be found at \href{https://github.com/CMBSPEC/SZpack}{github.com/CMBSPEC/SZpack}}. This new version contains a new fully-integrated {\tt Python} wrapper and the extension of several so far unreleased modules, including the code necessary to calculate the effects presented in this work and beyond.

This paper is arranged as follows, in Section~\ref{sec:Theory} we present the analytic forms of the SZ anisotropic scattering kernels and the mathematical background behind them. In Section~\ref{sec:cmb_anis}, we discuss the impact of anisotropies in the photon distribution and Section~\ref{sec:e_anis} examines the effects of anisotropies in the electron distribution. Section~\ref{sec:hEnT} uses a range of simple models of non-thermal electron momentum distributions to explore the effects of these components as both isotropic and anisotropic contributions. In Section~\ref{sec:discussion}, we discuss the impacts of all these contributions to the SZ effects and we conclude in Section~\ref{sec:conclusion}.

%--------------------------------------------------------------------------------
\section{Theory}
\label{sec:Theory}
%--------------------------------------------------------------------------------
In considerations of the SZ effect, an assumption of isotropy in both the photon and electron populations is usually taken. 
To define anisotropic scattering processes we must first determine what exactly is meant by the anisotropic scattering kernel. While in principle it would be possible to consider both an anisotropic electron and photon field, in this work we will only consider each separately. It should be noted that this paper focuses on the SZ scattering and thus assumes everything is in the Doppler-dominated regime where the electron energy greatly exceeds the photon energy and Klein-Nishina corrections can still be neglected.

\begin{figure} 
    \centering
    \includegraphics[width=0.95\columnwidth]{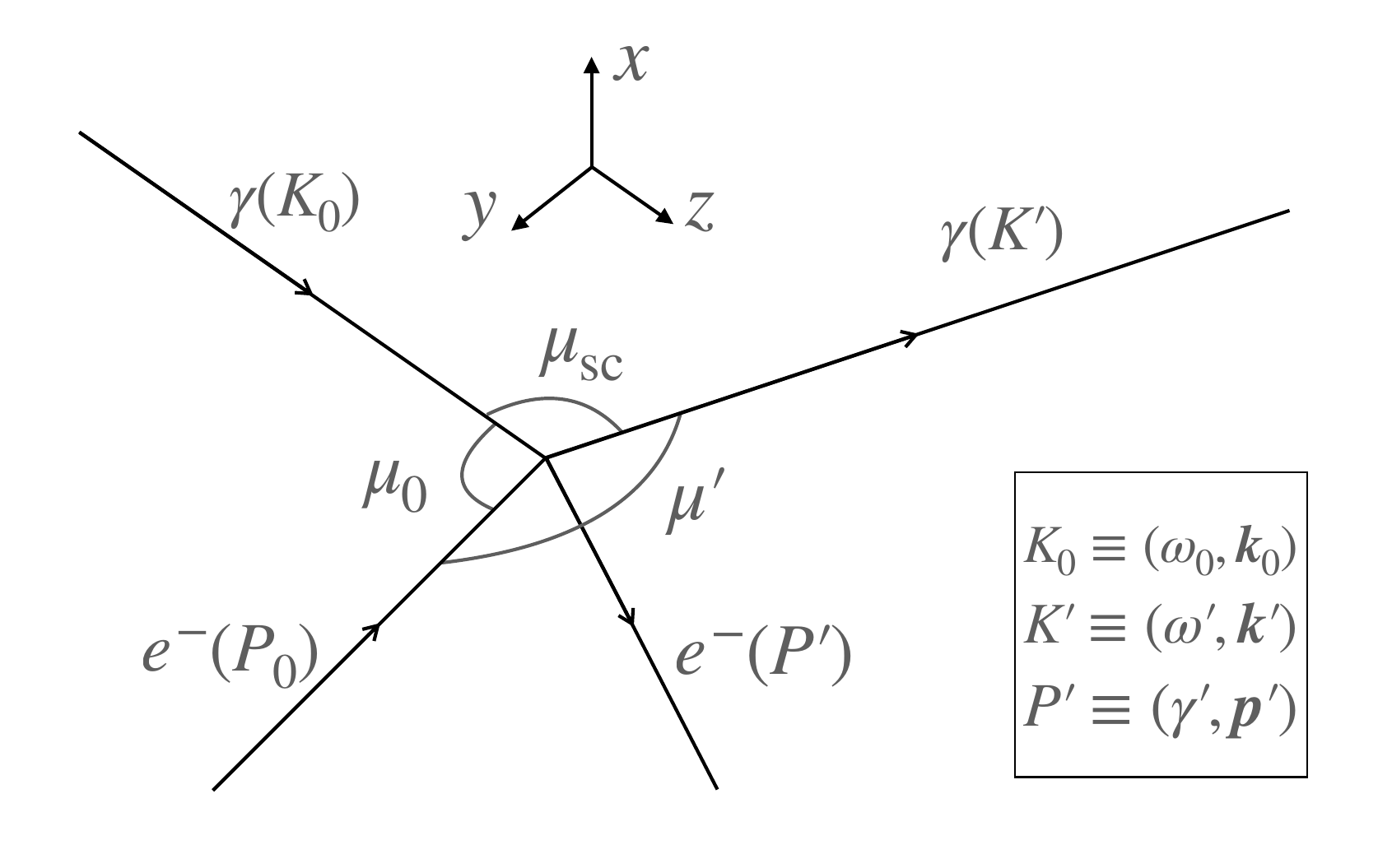}
    \caption[Compton Scattering Schematic]{Here $\Kz$ and $\Kp$ are the four-vectors for the incoming and outgoing photons, with dimensionless frequencies $\omegaz$ and $\omegap$ and dimensionless momenta $\kzb$ and $\kpb$. $\Pz$ and $\Pp$ are the four vectors for the incoming and outgoing electrons, with dimensionless energies $\gammaz$ and $\gammap$ and momenta $\pzb$ and $\ppb$. Then $\muz$, $\mup$ and $\musc$ are the cosines of the marked angles. Note that since the scattering is three-dimensional the relationship between these angles is complex. The axes are to emphasise the dimensionality of the scattering, with $z$ aligned along the direction of the incoming photon.}
    \label{int:fig:CS_schematic}
\end{figure}

To begin, it is convenient to define the relevant quantities involved in the SZ scattering process. The process is that of an incoming photon and electron, exchanging energy and momentum. The interaction takes the form $\gamma(\Kz)+e^{-}(\Pz)\leftrightarrow\gamma(\Kp)+e^{-}(\Pp)$ where here, $\Kz$, $\Pz$, $\Kp$ and $\Pp$ are the associated 4-vectors of the involved particles. A schematic of this process can be seen in Fig.~\ref{int:fig:CS_schematic}.
Then $n(\kzb)$ is the photon occupation number at dimensionless frequency $\omegaz=h\nu_0/\me c^2$, and $f(\pzb)$ is the electron momentum distribution function; $\pzb$, $\ppb$, $\kzb$ and $\kpb$ are the three-vectors associated with $\Pz$, $\Pp$, $\Kz$ and $\Kp$ respectively. We use energies and momenta in units of $\me c^2$ and $\me c$, respectively, such that $\omegaz$ and $\omegap$, the dimensionless frequencies, denote also the photon energies; $\gammaz$ and $\gammap$ are the electron energies and also the Lorentz factors, i.e., $\gammaz=E_0/\me c^2 = \sqrt{1+\pz^2} = 1/\sqrt{1-\betaz^2}$, where $\pz$ is the dimensionless momentum, and $\betaz=\varv/c=\gammaz\pz$ the dimensionless speed.
It is also useful to define three angles via their cosines, such that $\muz = \pzbh\cdot\kzbh$, $\mup = \pzbh\cdot\kpbh$ and $\musc = \kzbh\cdot\kpbh$ where the `hats' indicate that the vectors are normalised.

%--------------------------------------------------------------------------------
%--------------------------------------------------------------------------------

While in principle an anisotropic distribution can be expressed in any frame, for anisotropic SZ considerations there is a preferred frame for the spherical harmonic decomposition of the anisotropy. In an expansion about the axis of the photon $\Kz$ (i.e., the $z$-axis, as is depicted in Fig.~\ref{int:fig:CS_schematic}), an anisotropic photon distribution can be expressed\footnote{This now uses $x'=h\nu'/\kB\Tcmb$ as dimensionless frequency.} as $n(\kpb) = \sum_{\ell,m} n_{\ell m}(x') Y_{\ell}^m(\musc, \phi_{\rm sc})$. However, as will be justified in Appendix~\ref{app:derivation}, in this frame, as long as either the photon or electron distribution is isotropic, only the $m=0$ component will contribute to the distortion. This means the scattering of {\it any} anisotropic photon distribution by an isotropic electron distribution can be obtained by inserting $n(\kpb) = \sum_\ell n_\ell(x')P_\ell(\musc)$ instead, where $P_\ell$ denote the Legendre polynomials. Here the Legendre transform is defined as $n_\ell(x) = \frac{2\ell+1}{2}\int_{-1}^1 \D\mu  P_\ell(\mu) \,n(x, \mu)$, with $n(x, \mu)=\int \frac{\D\phi}{2\pi}n(x,\mu, \phi)$.
Similarly, the scattering of isotropic photons by an anisotropic electron distribution can be computed using $f(\pzb)=\sum_{\ell} f_{\ell}(\pz) P_{\ell}(\mu_0)$ with $\mu_0=\pzbh\cdot\kzbh$ and similarly for $f(\ppb)$. Note that $\ppb=\pzb+\kzb-\kpb$, such that the related dependencies can all be elimitated.

Let us start with the kinetic equation for the evolution of the photon at $\kzb$ in a general way as \citep[cf. e.g.,][]{Chluba2012}
%--------------------------------------------------------------------------------
\begin{align}
\label{hea:eqn:CNdef} 
\frac{\D{n(x_0, \kzbh)}}{\D{\tau}} &= \int \pz^2\D{\pz} \frac{\D{\mu_0}\D{\phi_0}\D{\musc}\D{\phi_{\rm sc}}}{\sigT\,\Ne}\,\frac{\D{\sigma}}{\D{\Omega}} \bigg\{f(\ppb) n(\kpb)[1+n(\kzb)] 
    - f(\pzb) n(\kzb)[1+n(\kpb)]\bigg\}\nonumber\\[2mm]
    &\simeq \int \pz^2\D{\pz} \frac{\D{\mu_0}\D{\phi_0}\D{\musc}\D{\phi_{\rm sc}}}{\sigT\,\Ne} \,\frac{\D{\sigma}}{\D{\Omega}} \bigg\{f(\ppb) n(\kpb) - f(\pzb) n(\kzb)\bigg\}
    \nonumber \\[2mm]
    &= -n(x_0, \kzbh)+\sum_{\ell=0}^{\infty}\sum_{m=-\ell}^{\ell}\int_0^\infty \D{t} \mathcal{P}_{\ell}^{m}(t) \,n_{\ell m}(t \,x_0),
\end{align} 
%--------------------------------------------------------------------------------
where $\D{\sigma}/\D{\Omega}$ is the differential Compton scattering cross section times the M{\o}ller speed\footnote{The normalization is $\int \frac{\D{\sigma}}{\D{\Omega}}\frac{\D{\mu_0}\D{\phi_0}\D{\musc}\D{\phi_{\rm sc}}}{4\pi\,\sigT}=1$.} \citep[see Eq.~(2) of][]{Chluba2012} and $\Ne$ the electron number density. 
Here, the first approximation ignores stimulated scattering effects (as is standard in SZ calculations) and the second equality uses $t\equiv \omegap/\omegaz$ to describe the change of the scattered photon energy. We also introduced the scattering optical depth along the photon path by\footnote{In this definition, $l$ should be thought of as an affine parameter along the photon worldline which transforms like time under Lorentz boosts.} ${\rm d} \tau = \sigT \Ne {\rm d} l$.

The photon scattering kernel $\mathcal{P}^{m}_{\ell}(t)$ averaged over the electron distribution can then be obtained using (see Appendix~\ref{app:derivation} for details)
%--------------------------------------------------------------------------------
\begin{equation} 
\label{eq:def_gen}
\begin{split}
    \mathcal{P}_{\ell}^{m}(t) &= \sum_{\ell'=0}^{\infty}\sum_{m'=-\ell'}^{\ell'}\int_{p_{\rm min}(t)}^\infty  \frac{\pz^2 \D{\pz}  f_{\ell'm'}(p_0)}{\Ne} \,K^{m m'}_{\ell\ell'}(t,\pz),
    \\[2mm]
    K_{\ell\ell'}^{m m'}(t,\pz) &\equiv \int \frac{\D{\mu_0}\D{\musc}\D{\phi_{\rm sc}}}{\sigT} \frac{\D{\phi_0}}{\D{t}} \frac{\D{\sigma}}{\D{\Omega}} Y_\ell^m(\musc, \phi_{\rm sc}) Y_{\ell'}^{m'}(\mu_0, \phi_0),
\end{split} \end{equation}
%--------------------------------------------------------------------------------
with $K_{\ell\ell'}^{m m'}(t,\pz)$ denoting the general multipole scattering kernel, and where we have used $p_{\rm min}(t) = \sinh(|\log(t)|/2)$. 

However, given the comments above it is sufficient to compute the individual photon, $K^{\gamma}_\ell=K_{\ell0}^{00}/\sqrt{2\ell +1}$, and electron, $K^{\rm e}_{\ell}=K_{0\ell}^{00}/\sqrt{2\ell +1}$, scattering kernels as (see Appendix~\ref{app:derivation} for details)
%--------------------------------------------------------------------------------
\begin{equation} \begin{split}
    K^{\gamma}_{\ell}(t,\pz) &\equiv \int \frac{\D{\muz}\D{\musc}\D{\phi_{\rm sc}}}{4\pi\,\sigT} \frac{\D{\phi_0}}{\D{t}} \frac{\D{\sigma}}{\D{\Omega}} P_\ell(\musc)\\
    K^{\rm e}_{\ell}(t,\pz) &\equiv \int  \frac{\D{\muz}\D{\musc}\D{\phi_0}}{4\pi\,\sigT} \frac{\D{\phi_{\rm sc}}}{\D{t}}\frac{\D{\sigma}}{\D{\Omega}} P_\ell(\muz)
\end{split} \end{equation}
%--------------------------------------------------------------------------------
and then average over an isotropic incoming electron or photon distribution respectively. 
Using the relations $n_{\ell 0}=\sqrt{4\pi / (2\ell+1)}\,n_\ell$ and $f_{\ell 0}=\sqrt{4\pi/ (2\ell+1)}\,f_\ell$, we then have two types of scattering problems, one for anisotropic photons scattered by isotropic electrons:
%--------------------------------------------------------------------------------
\begin{subequations}
\label{hea:eqn:gamma_aniso}     
\begin{align} 
\frac{\D{n(x_0, \kzbh)}}{\D{\tau}} &= -n(x_0, \kzbh)+\sum_{\ell=0}^{\infty}\int_{0}^{\infty} \D{t} \mathcal{P}^\gamma_{\ell}(t)\,n_{\ell}(t \,x_0)
\\
\mathcal{P}^\gamma_{\ell}(t)
&=
 \int_{p_{\rm min}(t)}^\infty  \frac{4\pi \pz^2 \D{\pz}  f_0(p_0)}{\Ne} \,K^{\gamma}_{\ell}(t,\pz),
\end{align}
\end{subequations}
%-----------------------------------------
the other for anisotropic electrons scattering isotropic photons:
%-----------------------------------------
\begin{subequations}
\label{hea:eqn:e_aniso}     
\begin{align} 
\frac{\D{n(x_0, \kzbh)}}{\D{\tau}} &= -n(x_0, \kzbh)+\int_{0}^{\infty} \D{t} \mathcal{P}^{\rm e}_0(t) \,n_0(t \,x_0),
\\
\mathcal{P}^{\rm e}_0(t) &= \sum_{\ell=0}^{\infty}\int_{p_{\rm min}(t)}^\infty  \frac{4\pi \pz^2 \D{\pz}  f_{\ell}(p_0)}{\Ne} \,K^{\rm e}_{\ell}(t,\pz). 
\end{align}
\end{subequations}
%--------------------------------------------------------------------------------
We note here that in cases where anisotropic electron distributions scatter isotropic photons, the scattered photon field becomes anisotropic. Similarly, anisotropic photons scattered by isotropic electrons generate anisotropies in the electron distribution, although this effect is negligible for the SZ regime. Therefore, the description given above can only be applied in the optically thin limit as relevant to the SZ effect since otherwise one would have to allow for both electron and photon anisotropies to evolve. 

%--------------------------------------------------------------------------------
%\subsection{Isotropic scattering kernel}
%\label{sec:Anis_photons}
%--------------------------------------------------------------------------------
In \cite{Ensslin2000}, kernel $K^\gamma_{0}(t,\pz)\equiv K^{00}_{00}(t,\pz)$ for isotropic media was given and can be expressed as 
%--------------------------------------------------------------------------------
\begin{equation} \begin{split}
    K^\gamma_{0}(t,\pz) &= \frac{3}{32\pz^6 t}\Bigg\{2t(1+t)(3+2\pz^2)\left[|\log(t)|-2{\rm sinh}^{-1}(\pz)\right]\\
    &\qqquad-|1-t|\left[1+(10+8\pz^2 +4\pz^4)t+t^2\right]
    %\\&\qqquad
    +4t(1+t)\frac{\pz(3+3\pz^2+\pz^4)}{\sqrt{1+\pz^2}} \Bigg\}.
\end{split} \end{equation}
%--------------------------------------------------------------------------------
This also directly determines $K^{\rm e}_{0}(t,\pz)$.
The expressions for other cases are, however, generally speaking complicated to calculate -- and in such circumstances where anisotropy in both populations exist \citep[beyond those of, for instance, a Doppler shift between frames that can be accounted for with a correction to the measured angles and frequencies directly as is described in e.g.,][ henceforth CNSN]{Chluba2012}, 
it may be simplest to calculate the entire collision integral numerically.

%--------------------------------------------------------------------------------
\subsection{Anisotropy in the photon population}
\label{sec:Anis_photons}
By considering anisotropy in each population individually, we can learn about the effects of each on the observed spectrums. Here, we first focus on the effects of an anisotropic photon distributions, with the effects of anisotropic electrons discussed in Section~\ref{sec:Anis_electrons}.

The process to determine the required kernels is detailed in Appendix~\ref{app:derivation}, and 
the kernels can be calculated to arbitrary $\ell$ where, for example,
%--------------------------------------------------------------------------------
\begin{align}
    K^{\gamma}_1(t,\pz) &= \frac{1}{32\pz^8 t}\Bigg\{-|1-t|\,\bigg[4\pz^6(1+t+t^2)-2\pz^4(1-14t+t^2)
    \nonumber\\&\qqquad  
    +\pz^2(31+166t+31t^2)+5(11+38t+11t^2)\bigg]
    \nonumber\\&\quad 
    +\frac{2\pz(1+t)}{\gamma_0}\bigg[2\pz^6(1-t+t^2)+30\pz^4t
    %\\[-1mm]&
    +\pz^2(11+142t+11t^2) +15(1+8t+t^2)\bigg]
    \\[-1mm]\nonumber& \qquad 
    +3(1+t) \bigg[4\pz^4 t+2\pz^2(1+17t+t^2)
    +5(1+8t+t^2)\bigg]
    %\\[-1mm]&\qquad  
    \left[|\mathrm{log}(t)|-2\mathrm{sinh}^{-1}(\pz)\right] \Bigg\},
\end{align}
defines the redistribution of photons from the dipolar anisotropy. Similarly, the scattering of photons from the quadrupolar anisotropy is given by
\begin{align}
    K^{\gamma}_2(t,\pz) &= \frac{3}{1280\pz^{10} t^2}\Bigg\{ -|1-t|\,\bigg[32\pz^8(1+t+t^2+t^3+t^4)
    %\\&\qquad  
    -16\pz^6(1+t-24t^2+t^3+t^4)
    \nonumber\\&\qquad  
    +4\pz^4(3+278t+1478t^2+278t^3+3t^4)
    %\\&\qquad  
    +30\pz^2(1+206t+666t^2+206t^3+t^4)
    \nonumber\\&\qquad  
    +35(3+178t+478t^2+178t^3+3t^4)\bigg]
    \nonumber\\[-1mm]&
    \quad+\frac{8\pz(1+t)}{\gamma_0}\bigg[4\pz^8(1-t+t^2-t^3+t^4)+120\pz^6t^2
    %\\&\qquad  
    +20\pz^4t(7+80t+7t^2)
    \\\nonumber&\qquad  
    +525t(1+5t+t^2)
    %\\&\qquad  
    +25\pz^2t(25+161t+25t^2)\bigg]
    \\[-1mm]\nonumber& 
    \quad+20t(1+t)\bigg[8\pz^6 t+12\pz^4(1+15t+t^2)
    \\[-1mm]\nonumber&\qquad  
    +90\pz^2(1+7t+t^2)+105(1+5t+t^2)\bigg]
    %\\[-1mm]&\qquad  
    \left[|\mathrm{log}(t)|-2\mathrm{sinh}^{-1}(\pz)\right] \Bigg\}.
\end{align}
%--------------------------------------------------------------------------------
It is evident that these quickly increase in complexity, while still maintaining a fairly consistent structure. In particular, there always arise polynomial coefficients for three parts of the equation -- a part proportional to $|\log(t)|-2\sinh^{-1}(\pz)$; a part proportional to $|1-t|$; and a purely polynomial part with a leading factor of $\pz(1+t)/\gamma_0$.

This continues to be true throughout all these kernels for photon anisotropy (and indeed for electron anisotropy). By expanding the legendre polynomials as $P_\ell(\musc) = \sum_{k=0}^\ell\binom{\ell}{k}\binom{\ell+k}{k} \left(\frac{\musc-1}{2}\right)^k$, a general form can be found, comprising of two main components,
%--------------------------------------------------------------------------------
\begin{align} 
\label{eq:K_gamma_gen_res}
    K^{\gamma}_\ell(t,\pz) =& \sum_{k=0}^\ell\binom{\ell}{k}\binom{\ell+k}{k}\left[\frac{3}{8\pz} 
    (G_k+H_k)\right]; 
    \nonumber \\[2mm]
    G_k =& \sum_{n=0}^{k}\frac{(-1)^n}{2n+1} \binom{k}{n}\,\frac{|1-t|^{2(k-n)}}{\gammaz (4t)^{k}}
    \left((1+t)^{2n+1}-\left[\frac{\gammaz|1-t|}{\pz}\right]^{2n+1}\right), 
    \nonumber \\[2mm]
    H_k =& \frac{(1+t)}{4 t \pz^{2k+5}}\Bigg[5\gamma_0^2 (1+t)^2  h_0^{(k)} 
    -\left[12 \gammaz^2 t+ (5+2\pz^2)(1+t)^2\right]h_2^{(k)} 
    +4t\,(3+2\pz^2)h_4^{(k)}\Bigg].
\end{align}
%--------------------------------------------------------------------------------
It is immediately clear that $G_k$ is a purely polynomial function and $H_k$ depends on the generator functions $h_m^{(k)}$ (see Appendix~\ref{app:derivation}), which have been defined to be the integrals 
\begin{equation} 
\label{eq:def_h}
\begin{split}
    h_m^{(k)} &= \int_{\gamma_0}^{\sqrt{1+\frac{|1-t|^2}{4t}}}\frac{\sqrt{y^2-1}\,(1-y^2)^k\,y^m}{y^6} \D{y};
\end{split} 
\end{equation}
with $y = \sqrt{1+\frac{1}{2}\pz^2(1-\musc)}$. These can be expressed analytically defining the first few terms specifically, using
%--------------------------------------------------------------------------------
\begin{equation} 
\begin{split}
    j^{(n)} &= \frac{\pz^{2n+1}}{\gammaz^5}-\frac{|1-t|^{2n+1}}{(4t)^{n-2}(1+t)^5}; \nonumber\\ \nonumber
    h_0^{(0)} &= -\frac{1}{15}\big[2j^{(2)}+5j^{(1)}\big], \;\; h_0^{(1)} = \frac{1}{5}\big[j^{(2)}\big], \;\;
    h_2^{(0)} = -\frac{1}{3}\big[j^{(2)}+j^{(1)}\big],  \\ \nonumber
    h_0^{(2)} &= \frac{1}{2} \Big[|\log(t)|-2\sinh^{-1}(\pz)\Big] +\frac{1}{15}\Big[23j^{(2)}+35j^{(1)}+15j^{(0)} \Big].
\end{split} 
\end{equation}
%--------------------------------------------------------------------------------
Then the higher $k$ terms can be expressed as
%--------------------------------------------------------------------------------
\begin{equation} 
\begin{split}
    h_0^{(k)} &= \frac{(2k+1)!!}{2^{k-2}(k-2)!} \left(\frac{h_0^{(2)}}{15}-\sum_{n=3}^{k}\frac{2^{n-3}(n-3)!}{(2n+1)!!}(-1)^n j^{(n)} \right)\; {\rm for\;} k\geq3;\\
    h_2^{(k)} &= \frac{-1}{2k+3}\left(5h_0^{(k+1)}+(-1)^k j^{(k+1)}\right) \qqquad\qquad\quad\; {\rm for\;} k\geq1; 
    \\ \nonumber
    h_4^{(k)} &= \frac{-1}{2k+3}\left(3h_2^{(k+1)}+(-1)^k \left[j^{(k+2)}+j^{(k+1)}\right]\right).
\end{split} 
\end{equation}
%--------------------------------------------------------------------------------
In general, we see the same components as before with the `log' part coming from the $h_0^{(2)}$ term which carries into all the higher order terms as well. $j^{(n)}$ clearly contains both a $|t-1|$ term and a polynomial part, which, with a consideration of $H_k$, will always have a leading term of $(1+t)$ as expected. $G_k$ directly contains both the $|t-1|$ and $(1+t)$ polynomial parts.

It should be noted that in the current form these analytic expressions are not numerically stable as $\ell$ increases and $t\rightarrow1$. An exploration of the numerical stability of both anisotropic kernels can be found in Appendix~\ref{app:numericalStability}.

\begin{figure}
    \centering
    \includegraphics[width=0.8\linewidth]{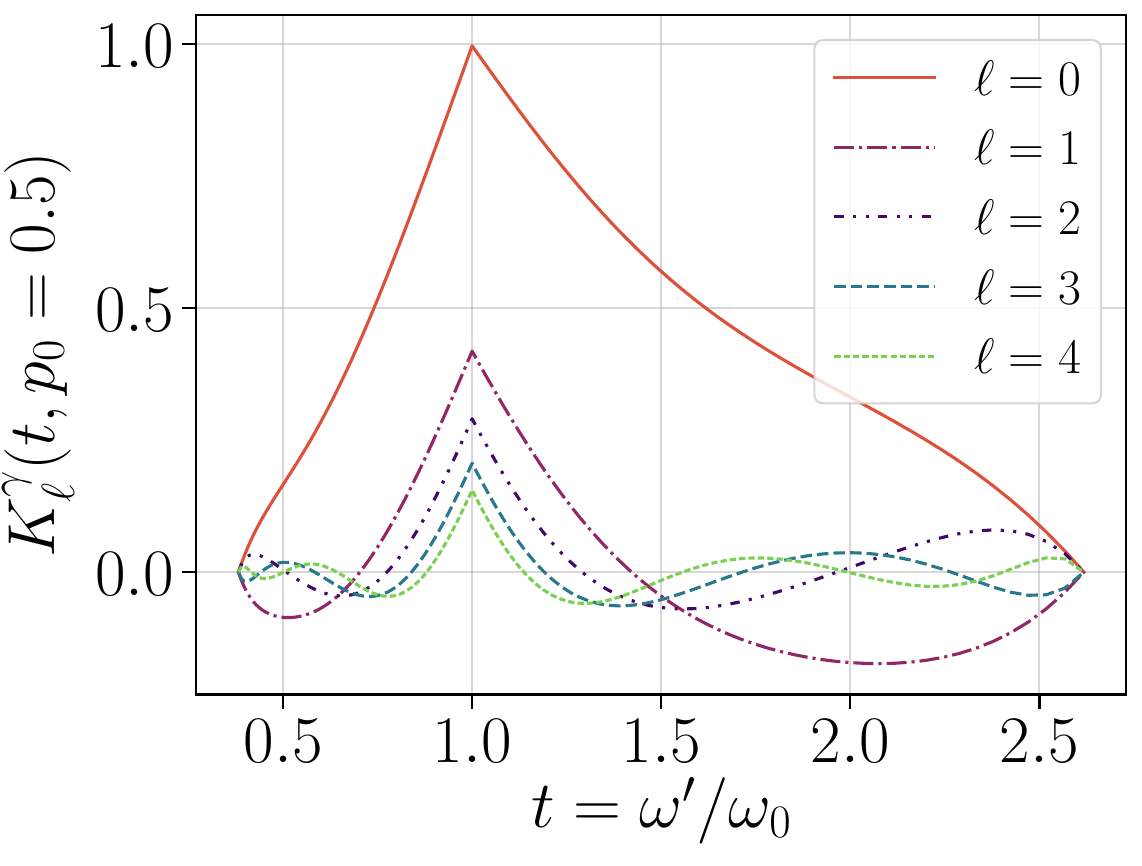}\\
    \caption[The anisotropic photon scattering kernel]{The first four multipoles, $\ell\leq4$, of the anisotropic photon multipole scattering kernel, $K^\gamma_\ell$. Here we have arbitrarily set $\pz=0.5$. Note that all the higher multipoles have negative components, indicating redistribution between multipoles. }
    \label{fig:1_K_photon_ani}
\end{figure}

The photon scattering kernels for the first four multipoles can be seen in Fig.~\ref{fig:1_K_photon_ani}. Here we see, as has been long established \citep[e.g.,][]{Sunyaev1980,Sazonov2000}, that the monopole scattering kernel has a cusp at $t=1$, driven mathematically by the absolute values of $|\log(t)|$ and $|t-1|$. From a physical standpoint, this is indicating that a photon is most likely (modally) to maintain its same energy and merely be deflected under a scattering. However, the tail to higher energies indicates, as expected, the propensity of SZ scatterings to move photons from lower energies to high energies -- that is, to upscatter them.

These kernels are consistent with those displayed in, for instance, \cite{Chluba2014b}, and have also been verified numerically.
The kernels for higher multipoles all maintain a cusp at $t=1$ (as can be predicted from the mathematical forms continuing to contain these absolute values). However, they all also contain negative sections, which are harder to interpret -- it must be remembered that each multipole follows the scattering caused by the population of photons in a certain multipole -- and as such, the negative sections indicate the scattering of photons out of the given multipole. That is, the SZ scattering causes an isotropisation of the photon distribution. Isotropising the distributions and driving them towards equilibrium in terms of the momentum distributions thus go hand in hand. Indeed, one cannot have thermal momentum distributions in the presence of anisotropies, as for example the superposition of blackbodies of different temperatures is not a blackbody \citep[e.g.,][]{Chluba2004, Stebbins2007}.

It is also interesting to note that each multipole has a kernel that crosses zero, $\ell$ times on either side of $t=1$. That is, we can consider for each multipole the kernel is composed of $2\ell+1$ regions where the kernel is alternately positive or negative. The large negative region in $\ell=1$ at higher values of $t$ leads to an effect opposing that of the monopole, that is, the dipole is scattered preferentially to lower frequencies. This in fact is also true for the $\ell=3$ and $\ell=4$ cases, while the $\ell=2$ scattering leads to net upscattering like the monopole.

It should, however, be noted, that the amplitude of these scatterings falls rapidly with increasing $\ell$ -- even supposing the amplitude of the anisotropic components of the photon distribution were the equal to the monopole, for a thermal electron distribution (that is, relativistic Maxwell-Boltzmann) and CMB photons, the amplitude is suppressed with increasing $\ell$. 
This can be determined analytically, by integrating over frequency (or $t$) and taking an expansion about $\pz=0$.  The monopole kernel obeys (at all $\pz$) $\int_{t_+}^{t_-} \D{t} K_0(t,\pz) \equiv 1$. In a similar way, we can find, for $\ell\leq4$,
%---------------------------------------------------------------------------------
\begin{equation} \begin{split}
    \int_{t_+}^{t_-}  \D{t} K^{\gamma}_0(t,\pz) &\equiv 1; \\
    \int_{t_+}^{t_-}  \D{t} K^{\gamma}_1(t,\pz) &\simeq -\frac{2}{15}\pz^2+\frac{4}{75}\pz^4 -\frac{8}{735}\pz^6-\frac{32}{2205}\pz^8;\\ %&\qquad+\mathcal{O}(\pz^{10});\\
    \int_{t_+}^{t_-}  \D{t} K^{\gamma}_2(t,\pz) &\simeq \frac{1}{10}-\frac{1}{5}\pz^2+\frac{48}{175}\pz^4-\frac{16}{49}\pz^6+\frac{160}{441}\pz^8; \\ %&\qquad+\mathcal{O}(\pz^{10}); \\
    \int_{t_+}^{t_-}  \D{t} K^{\gamma}_3(t,\pz) &\simeq \frac{2}{35}\pz^2-\frac{24}{175}\pz^4+\frac{32}{147}\pz^6-\frac{128}{441}\pz^8; \\ %\\&\qquad+\mathcal{O}(\pz^{10}); \\
    \int_{t_+}^{t_-}  \D{t} K^{\gamma}_4(t,\pz) &\simeq \frac{8}{315}\pz^4-\frac{32}{441}\pz^6+\frac{640}{4851}\pz^8. %+\mathcal{O}(\pz^{10}). \\
\end{split} \end{equation}
%--------------------------------------------------------------------------------
Note that the integration limits here are that $t_\pm = (\gammaz\pm\pz)/(\gammaz\mp\pz)$. One also naturally has $t_+=1/t_-$.
These expressions return the expected result obtained in the Thomson limit that only the $\ell=0$ and $2$ components contribute at zeroth order -- with the $\ell=2$ component at $1/10$ the amplitude of the monopole \citep[see also][]{Chluba2014b}.

%--------------------------------------------------------------------------------
%--------------------------------------------------------------------------------

For further comparisons, we also give the first moments of the kernels, describing the energy exchange between the electrons and photons:
%---------------------------------------------------------------------------------
\begin{equation} 
\label{eq:first_moment}
\begin{split}
    \int_{t_+}^{t_-} \D{t} (t-1) K^{\gamma}_0(t,\pz) &\equiv \frac{4}{3}\,p_0^2; \\
    \int_{t_+}^{t_-}  \D{t} (t-1) K^{\gamma}_1(t,\pz) &\simeq -\frac{8}{15}\pz^2-\frac{4}{75}\pz^4 +\frac{8}{735}\pz^6+\frac{32}{2205}\pz^8;\\ %&\qquad+\mathcal{O}(\pz^{10});\\
    \int_{t_+}^{t_-}  \D{t} (t-1) K^{\gamma}_2(t,\pz) &\simeq \frac{2}{15}\pz^2-\frac{4}{35}\pz^4+\frac{232}{1225}\pz^6-\frac{2656}{11025}\pz^8;\\ %&\qquad+\mathcal{O}(\pz^{10}); \\
    \int_{t_+}^{t_-}  \D{t} (t-1) K^{\gamma}_3(t,\pz) &\simeq -\frac{2}{35}\pz^2+\frac{24}{175}\pz^4-\frac{32}{147}\pz^6+\frac{128}{441}\pz^8;\\ %&\qquad+\mathcal{O}(\pz^{10}); \\
    \int_{t_+}^{t_-}  \D{t} (t-1) K^{\gamma}_4(t,\pz) &\simeq -\frac{8}{315}\pz^4+\frac{32}{441}\pz^6-\frac{640}{4851}\pz^8. %+\mathcal{O}(\pz^{10}). \\
\end{split} \end{equation}
%--------------------------------------------------------------------------------
These values further confirm some of the statements made above. We also mention that by carrying out the thermal averages of these moments over the electron distribution we can reproduce the related moments given in the appendix of \citep{Chluba2012} used directly for the computations of the SZ signals with kinematic corrections.

We mention that for $\ell\leq 2$, \cite{Nozawa2013} provided analytical expressions for the photon scattering kernels that were then applied to moving electrons. However, we were unable to establish the direct link to our expressions. We confirmed our expressions numerically finding agreement with the numerical results for the kernels, providing a starting point for a future comparison.

%--------------------------------------------------------------------------------
\subsection{Anisotropy in the electron population}
\label{sec:Anis_electrons}
In a similar way the anisotropic electron multipole kernels can be derived (see Appendix~\ref{app:derivation_e} for details). For the monopole, one naturally has $K^{\rm e}_{0}=K^{\gamma}_{0}$. The dipole term then reads
%--------------------------------------------------------------------------------
\begin{subequations}
\label{eq:e_kernel_12}
\begin{align} 
    K^{\rm e}_{1}(t,\pz) &=
\frac{3}{32\gammaz\pz^7}\Bigg\{-\frac{|1-t|}{t}\,\bigg[-4 \pz^6 t^2 + 2\pz^4 t (5-t)
+ \pz^2(1+26 t+9 t^2)+(1+19 t+10t^2)\bigg]
\nonumber\\
    & \qquad   +2\gamma_0\pz  \bigg[2\pz^4 t(1-t)+2\pz^2(3+5 t)
    +3(3+6t+t^2)\bigg]
\\[-1mm]\nonumber
    &\qquad\quad   + \bigg[4\pz^4 (1+t)+2\pz^2(6+11 t+t^2)
    +3(3+6t+t^2)\bigg]
\left[|\mathrm{log}(t)|-2\mathrm{sinh}^{-1}(\pz)\right] \Bigg\},
\end{align} 
%--------------------------------------------------------------------------------
and
%--------------------------------------------------------------------------------
\begin{align} 
K^{\rm e}_{2}(t,\pz) &=
    \frac{3}{16\pz^8}\Bigg\{-\frac{|1-t|}{4t}\,\bigg[-8 \pz^6 t^2 (1-2t) + 8\pz^4 t (3-2t+t^2)
\nonumber\\
    &\qqquad\qquad 
    + 2\pz^2(1+45 t+27 t^2-t^3)+3(1+29 t+29t^2+t^3)\bigg]
\nonumber\\
    & +\frac{\pz}{\gamma_0}  \bigg[2\pz^6 t(1-3t+2t^2)+\pz^4 (7+18 t -9 t^2 +4 t^3)+6\pz^2(4+11 t+2t^2)
    +18(1+3t+t^2)\bigg]
\nonumber\\[-1mm]
    &\qquad\quad   + \bigg[2\pz^4 (1+t)+3\pz^2(3+8 t+t^2)
    +9(1+3t+t^2)\bigg]
\left[|\mathrm{log}(t)|-2\mathrm{sinh}^{-1}(\pz)\right] \Bigg\}
\end{align}
\end{subequations}
%--------------------------------------------------------------------------------
is found for the quadrupole.

Then, for general $\ell$, the kernel can be written as
%--------------------------------------------------------------------------------
\begin{align} 
\label{eq:e_kernel_ell}
    K^{\rm e}_{\ell}(t,\pz) &=\sum_{k=0}^{\ell}\binom{\ell}{k}\binom{\ell+k}{k}\Bigg[\frac{3}{32 \gammaz \pz^5} \Big(3 t^2 X_4^{(k)}-6\gammaz \, t(t+1) \, X_3^{(k)}+\left[(3+2\pz^2)(1+t^2) + 12 \gammaz^2 t\right] X_2^{(k)}
    \nonumber\\
    &\qqquad\qqquad\qqquad-2 \gammaz (3+2\pz^2)\,(1+t)\,X_1^{(k)}+(3+4\pz^2+4\pz^4)\,X_0^{(k)}\Big)\Bigg]; 
    \nonumber\\
    X_m^{(k)} &= \frac{(\gammaz-\pz)^{k+1-m}}{2^k \pz^{k+1}} \sum_{n=0}^{k}\,(-1)^{n}\,\binom{k}{n} \;g^{(n+1-m)};
\end{align}
%--------------------------------------------------------------------------------
Where we have defined the auxiliary function $g^{(\alpha)}$, which is given by
%---------------------------------------------------------
\begin{equation} 
\label{eq:e_kernel_gfunc}
\begin{split}
g^{(\alpha)} &=
\begin{cases}
-|\log(t)|+2\sinh^{-1}(\pz) &\alpha = 0
\\[2mm]
\displaystyle 
\frac{(t^\alpha+1)(t_+^\alpha-1)
    -{\rm sign}(t-1)(t^\alpha-1)(t_+^\alpha+1)}{2 \alpha}
    &{\rm otherwise;}
\end{cases} 
%\\
%t^\pm&=\frac{1\pm\betaz}{1\mp\betaz}.
\end{split} \end{equation}
%---------------------------------------------------------
Here the $|t-1|$ term has been split into a ${\rm sign}(t-1)(t^n-1)$ part, to keep the expression simpler. It is worth reiterating that little work has been done to study high order anisotropies in the microphysical electron distribution. A `dipole' would reflect the motion of a cell of electrons, as in the kinematic SZ correction, and quadrupoles may occur due to the presence of magnetic fields in clusters. However, higher multipoles of electron anisotropy remain broadly unexamined in SZ physics and are expected to be small due to rapid Coulomb scattering leading to isotropisation and thermalisation.

Once again, we note that these analytic forms are not numerically stable for arbitrary $\ell$, and a consideration of their numerical stability can be found in Appendix~\ref{app:numericalStability}.

%--------------------------------------------------------------------------------
\begin{figure}
    \centering
    \includegraphics[width=0.8\linewidth]{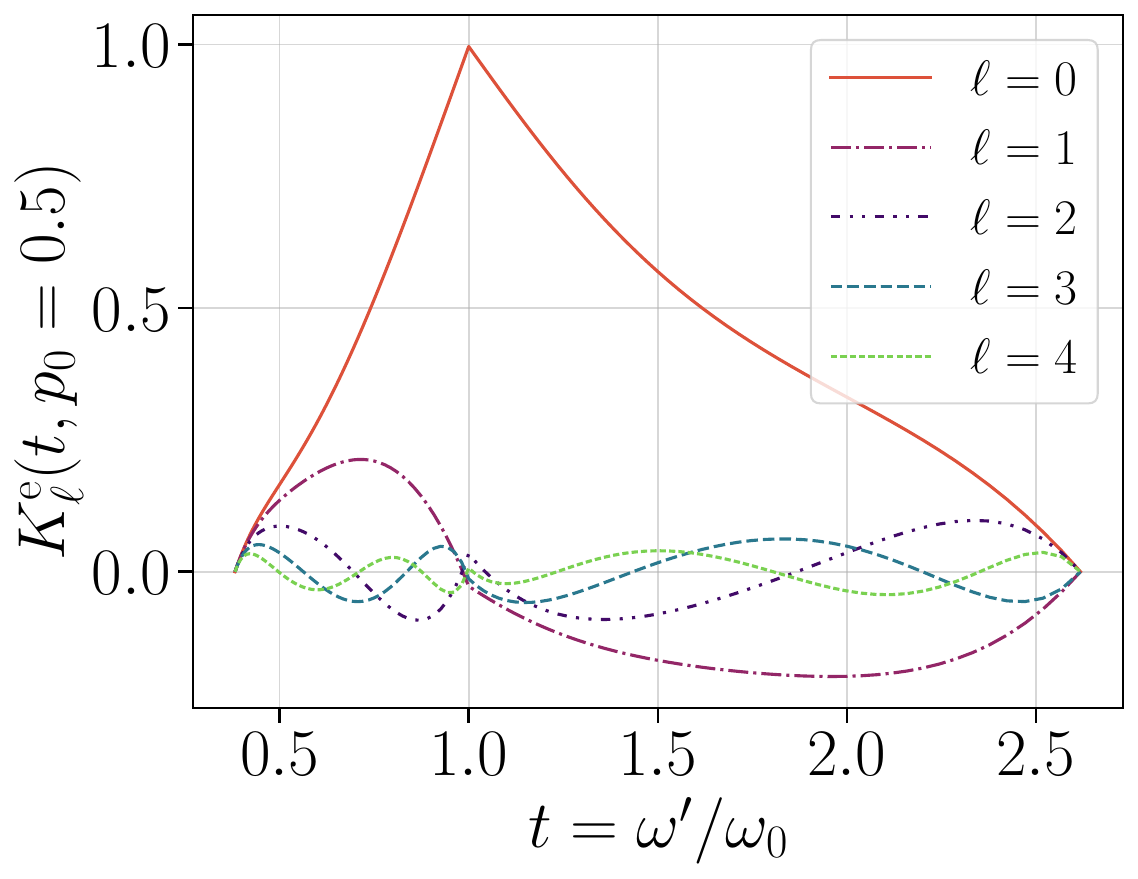}\\
    \caption[The anisotropic electron scattering kernel]{The first four multipoles, $\ell\leq4$, of the anisotropic electron multipole scattering kernel, $K^{\rm e}_{\ell}$. Here we have arbitrarily set $\pz=0.5$. Note that all the higher multipoles have negative components, indicating scattering between multipoles.}
    \label{fig:2_K_electron_ani}
\end{figure}
%--------------------------------------------------------------------------------

Nonetheless, the first four multipoles are displayed in Fig.~\ref{fig:2_K_electron_ani}. It is immediately clear that these are significantly harder to interpret than for the photon anisotropies. Firstly, it is evident that although there is a discontinuity in the gradient at $t=1$, there is no longer the clear cusp displayed in the photon kernels. Secondly, as has been long established, the dipole term relates to bulk motion, while the higher multipoles lead to significantly more complicated behaviour. Once again, when $t>1$ each multipole crosses zero, $\ell$ times. However, the distinct asymmetry in the kernels means that for the odd multipoles  for $t<1$ there are only $\ell-1$ crossings, while the even scattering kernels, have a more `cuspy' behaviour, and cross $\ell$ times.

Here also, the normalizations of the kernels are not the same as for the photon case. Analytically this can be explored by expanding as $\pz\rightarrow0$, so that
%--------------------------------------------------------------------------------
\begin{align}
    \int_{t_+}^{t_-}  \D{t} K^{\rm e}_{0}(t,\pz) &\equiv 1; \nonumber \\ \nonumber
    \int_{t_+}^{t_-}  \D{t} K^{\rm e}_{1}(t,\pz) &\simeq -\frac{1}{3}\pz+\frac{1}{6}\pz^3-\frac{1}{8}\pz^5+\frac{5}{48}\pz^7-\frac{35}{384}\pz^9; %+\mathcal{O}(\pz^{10});
\end{align} 
%--------------------------------------------------------------------------------
and $\int_{t_+}^{t_-}  \D{t} K^{\rm e}_{\ell>1}(t,\pz)=0$. This indicates that scattering of photons by electrons in anisotropies does not change the number of photons for $\ell>1$ and only leads to energy exchange. For the corresponding first moments, we find
%--------------------------------------------------------------------------------
\begin{align}
    \int_{t_+}^{t_-}  \D{t} (t-1) K^{\rm e}_{0}(t,\pz) &\equiv \frac{4}{3}p_0^2;
    \\ \nonumber
    \int_{t_+}^{t_-}  \D{t} (t-1) K^{\rm e}_{1}(t,\pz) &\simeq -\frac{1}{3}\pz-\frac{1}{2}\pz^3+\frac{5}{24}\pz^5-\frac{7}{48}\pz^7+\frac{15}{128}\pz^9;
    \\ \nonumber
    \int_{t_+}^{t_-}  \D{t} (t-1) K^{\rm e}_{2}(t,\pz) &\equiv \frac{2}{15}\pz^2;
\end{align} 
%--------------------------------------------------------------------------------
and $\int_{t_+}^{t_-}  \D{t} K^{\rm e}_{\ell>2}(t,\pz)=0$. This shows that only the first three electron multipoles leads to energy exchange with the photons.

%--------------------------------------------------------------------------------
\section{Scattering anisotropies in the CMB}
\label{sec:cmb_anis}
%--------------------------------------------------------------------------------
The most commonly considered anisotropies in SZ scattering are those caused by the intrinsic CMB fluctuations -- in particular, the polarised effects generated by the CMB quadrupole lead to fluctuations around $10^{-8}$ of the CMB temperature \citep[i.e.,][]{Sazonov1999}. Other anisotropies considered for the pSZ effect lead to signals of similar order of magnitude -- that is, from multiple scatterings within clusters or the higher order kSZ induced pSZ signals \citep[see][for overview]{Mroczkowski2019}. Specific anisotropies within the electron population will be discussed further in the next section. However, those that we mention here all focus on the polarised components rather than the general intensity, although the amplitudes of the signals are highly related. Moreover, in the literature, these predictions are generally calculated using simplistic models to generate the spectral dependence, instead of using the full cross-section. The kernels given here should thus allow for more general treatments.

%--------------------------------------------------------------------------------
\begin{figure}
    \centering
    \includegraphics[width=0.8\linewidth]{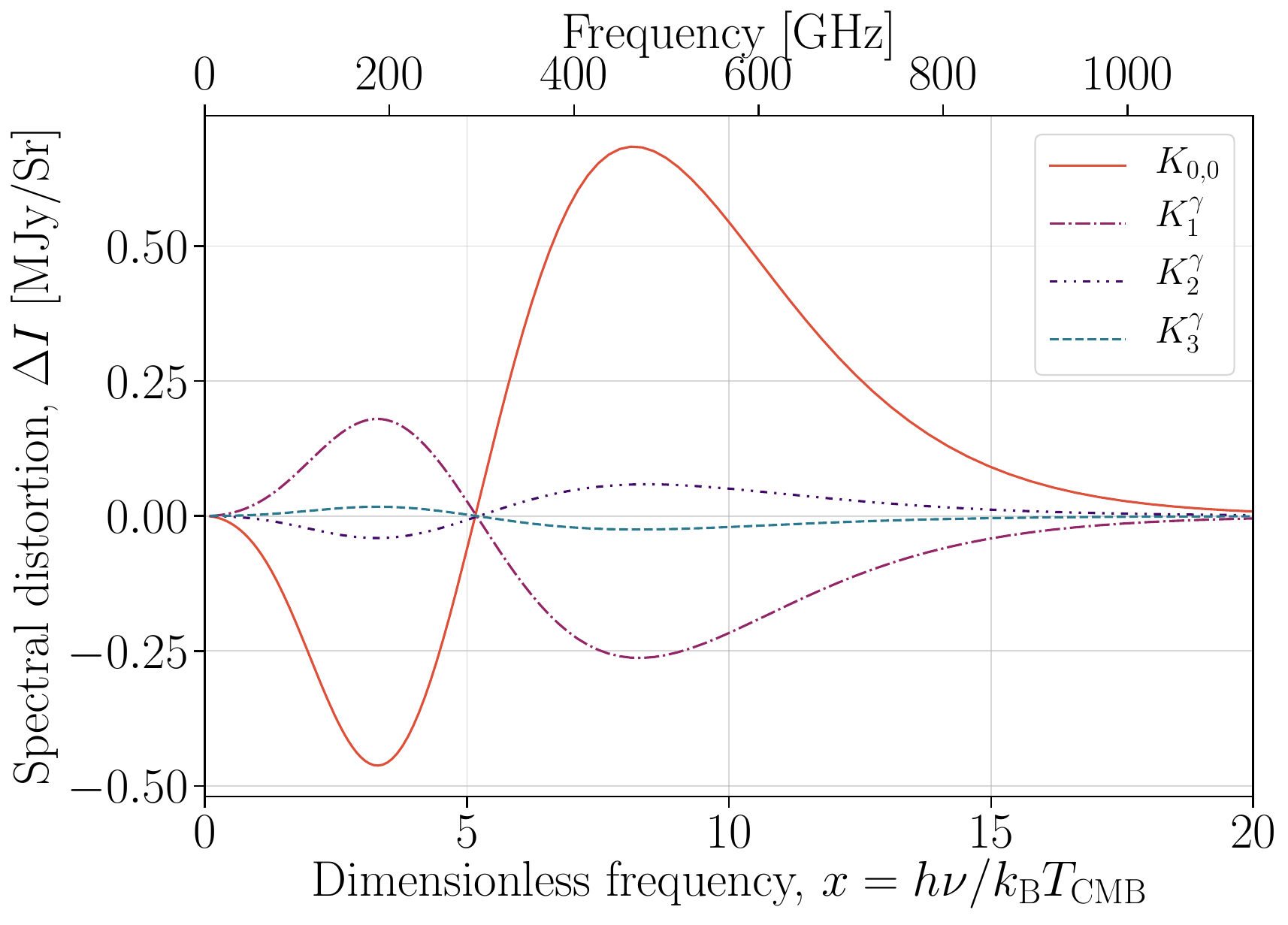}
    \caption[Anisotropic CMB SZ signals]{The SZ signals induced by anisotropies in the CMB. Here, the $K_{\ell}$ indicate which kernel is used. These are the signals caused by a 5~keV electron population and the first derivative of the CMB blackbody photon distribution as described in Eq.~\eqref{eqn:cmb_anis}. These signals use $y=10^{-4}$.}
    \label{fig:3_dist_photon_anis}
\end{figure}
%--------------------------------------------------------------------------------

To gain a heuristic understanding of the effects of the CMB anisotropies on the SZ effect, it is first worth detailing the form that these anisotropies will take. In particular, it is easy to show that temperature anisotropies in the CMB can be expressed as
%--------------------------------------------------------------------------------
\begin{equation} \label{eqn:cmb_anis} \begin{split}
    n_{\rm CMB}(\nu', T, \kpb) &= n(\nu, \Bar{T}, \kpb) +\left.\frac{\partial{n}}{\partial{T}}\right|_{\Bar{T}}(T(\kpb)-\Bar{T})+\mathcal{O}(\Delta T^2) \\
    &= n(x', \kpb) -x' \partial_{x'}{n}\left(\frac{T(\kpb)-\Bar{T}}{\Bar{T}}\right)+\mathcal{O}(\Delta T^2),
\end{split} \end{equation}
%--------------------------------------------------------------------------------
with $\Delta T=T(\kpb)-\Bar{T}$ and $\Bar{T}$ the average CMB temperature. We can also recall that $n(x)=1/(\expf{x}-1)$, which is the spectral shape relevant to the scattering of the CMB monopole. As such we can write the anisotropies, by expanding $[T(\kpb)-\Bar{T}]/\Bar{T}$ in terms of spherical harmonics, as discussed before. We neglected higher order temperature terms, given that these will be $\simeq 10^{-8}$ (with a $y$-type distortion spectrum), although for a consideration of polarization effects, this might be of interest.

Now it is clear that the spectral dependence of the higher moments is caused by the first derivative of the CMB blackbody -- these have been plotted in Fig.~\ref{fig:3_dist_photon_anis}. It should be noted that these have all been plotted with the same amplitude, i.e., assuming that $n_\ell(x)=-x\partial_x n=x\expf{x}/(\expf{x}-1)^2$, without the amplitude scaling from the $\Delta T/\Bar{T}$ expansion. That is, this figure shows the shapes of each component, but not the relative amplitudes for realistic CMB anisotropies. This figure also uses a fixed 5~keV electron population (following a relativistic Maxwell-Boltzmann distribution).

The first aspect to note, is that the intrinsic amplitude of the scattering signal from increasing multipoles decreases -- as may be expected from the decreasing amplitudes seen in the kernels in Fig.~\ref{fig:1_K_photon_ani}. Second, it is useful to note that the multipoles have alternating sign, with the even multipoles contributing a decrement at lower frequencies and an increase at higher frequencies, and the odd multipoles counteracting this behaviour. It is also worth emphasising that the null of these distributions do not lie at around 217~GHz as is seen for the primary CMB SZ distortion, but instead occurs at around 293~GHz even if not as visible in Fig.~\ref{fig:3_dist_photon_anis}. Given that the amplitude of the primary CMB anisotropies is $\Delta T/T\simeq 10^{-4}$, it is clear that the related scattering signals require very high sensitivity to be distinguished. However, this could for example open a novel way to test the redshift evolution of the lowest multipoles and possibly even the primordial dipole.

%--------------------------------------------------------------------------------
\subsection{Modelling kSZ effects using anisotropic scattering}
\label{sec:kSZ_photons}
While the photon kernel formalism is best adapted for intrinsic anisotropies present within the CMB, it is also possible to express the kinematic effects in terms of photon anisotropies, as is discussed in CNSN. This method allows for intrinsic CMB anisotropies to be easily combined with the kinematic corrections, but relies on approximating the kinematic effects in the limit of a small cluster speed $\betac=\varv_{\rm c}/c$. With the anisotropic scattering kernel approach given here, we can in principle extend to higher orders in $\betac$ without using a perturbative expansion in the energy exchange terms, thereby avoiding limitations from an asymptotic expansion (see CNSN for related discussion).

In particular, CNSN developed a framework wherein the signal is calculated in the cluster frame, where the radiation is then anisotropic due to the kinematic corrections. These anisotropies are calculated by means of a Lorentz boost of the photon number density and curtailed to order $\betac^2$, generating terms which are sums of terms of the form $x^k\partial_x^k n$ [see e.g., Eqs.~(11) and (12) of their paper]. The scattering problem in the cluster frame then boils down to a calculation of the photon monopole, dipole and quadrupole anisotropies. Once the scattering of the anisotropic photon distribution is computed in the cluster frame, it is then possible to transform back into the CMB (or observer) frame by a transformation of the relevant quantities -- that is, by expressing $x$ and the direction cosine of the cluster motion, $\muc$, with consideration to which frame they are being measured in and in which frame they are being used for SZ calculations [see, e.g., CNSN Eq.~(24)].
The benefit of this procedure is that it allows cleanly defining the scattering optical depth and temperature of the cluster in the rest frame of the moving cloud, thereby avoiding ambiguities.

For the purposes of this paper, it is sufficient here to merely describe that this transformation exists -- although it will be discussed further in Section~\ref{sec:kSZ_electrons}. It is worth emphasising, however, that a formulation like this allows for the combination of intrinsic CMB anisotropies with the kinematic SZ effects, although for the fullest precision would require the CMB anisotropies to be boosted themselves into the cluster frame by applying the aberration kernel \citep[e.g.,][]{Chluba2011abb, Dai2014abb}. However, these kinematic corrections to the primordial CMB anisotropies will be significantly smaller than the effects of the anisotropies themselves (as, for instance, the first correction will be multiplied by a factor of $\betac$). We thus do not consider this problem any further.

%--------------------------------------------------------------------------------
\section{Effect of electron anisotropies}
\label{sec:e_anis}
%--------------------------------------------------------------------------------
Anisotropic electron distributions lead to distinct SZ signals -- and can generally be expected to be caused by shocks or jets in clusters that lead to local anisotropies in the electron distribution. This has been examined in the case of polarised SZ effects in e.g., \cite{Khabibullin2018}, where they detailed how the magnetic field-induced electron anisotropy in shocks could generate polarised signals of a similar scale to those caused by photon anisotropy.

\begin{figure}
    \centering
    \includegraphics[width=0.8\linewidth]{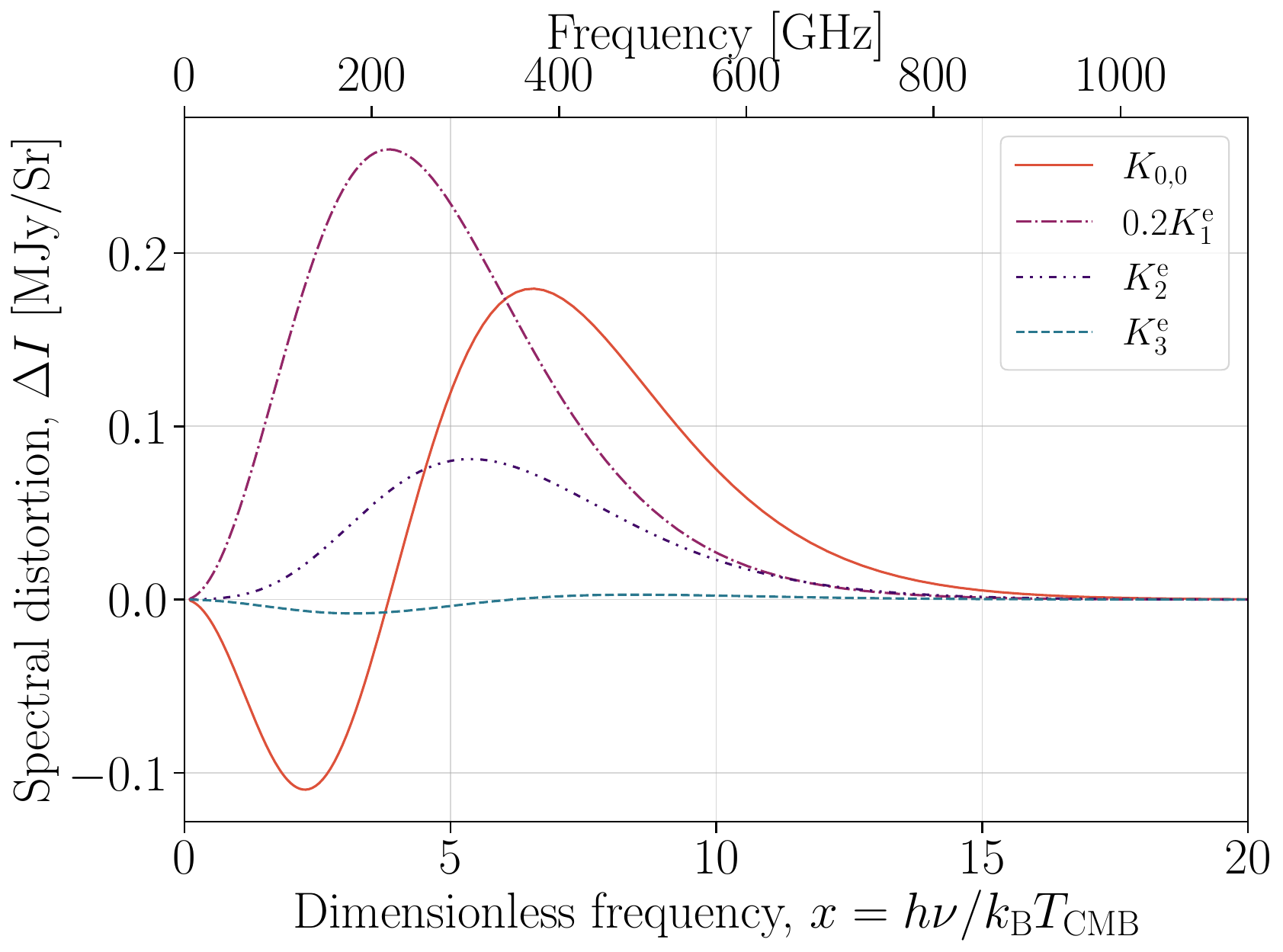}
    \caption[Anisotropic electron SZ signals]{The SZ signals induced by anisotropies in the electron distributions. Here, the $K_{\ell}$ indicate which kernel is used. We used a fixed electron momentum of $p_0 = 0.1$ to scatter the isotropic CMB blackbody spectrum. All of these signals are for $y=10^{-4}$ and $\tau = y/(\betaz^2/3)$. We set $p^2 f_\ell(p)=1$ for all curves but scaled the curve for $\ell=1$ by $0.2$ to make it more comparable to the other signals.}
    \label{fig:4_dist_electron_anis}
\end{figure}

In general, we find that the signals created by anisotropies in the electron population are far more akin to the kinematic SZ signals in shape and behaviour -- and we will examine the kinematic case later in this section. However, to first order this  can be seen by considering the effects of a monoenergetic electron distribution at various multipoles scattering the isotropic CMB. This is displayed in Fig.~\ref{fig:4_dist_electron_anis} for high energy electrons [$\pz=0.1$ which is comparable to a temperature $\kB T_{\rm eff}=(\betaz^2/3)\; (\me c^2)\simeq1.7$~keV].
%--------------------------------------------------------------------------------
%--------------------------------------------------------------------------------

Here the $\ell=0$ case is close to the standard SZ signal. The signal created by the scattering of photons by electrons in the dipole anisotropy greatly resembles the kinematic dipole in shape, with $p_0>0$ implying motion towards (i.e., up-scattering of CMB photons) the observer. The quadrupole component is entirely positive, but generates a sizeable signal in this basic analysis. The octopole has a null at around 350~GHz, with a negative component at low frequencies and positive at high frequencies. Again, however, we do see a tendency towards decrease in amplitude with increasing multipoles (although the dipole term here greatly exceeds the monopole signal -- and for clarity has been suppressed by a factor of 0.2 within Fig.~\ref{fig:4_dist_electron_anis}).

\begin{figure}
    \centering
    \includegraphics[width=0.8\linewidth]{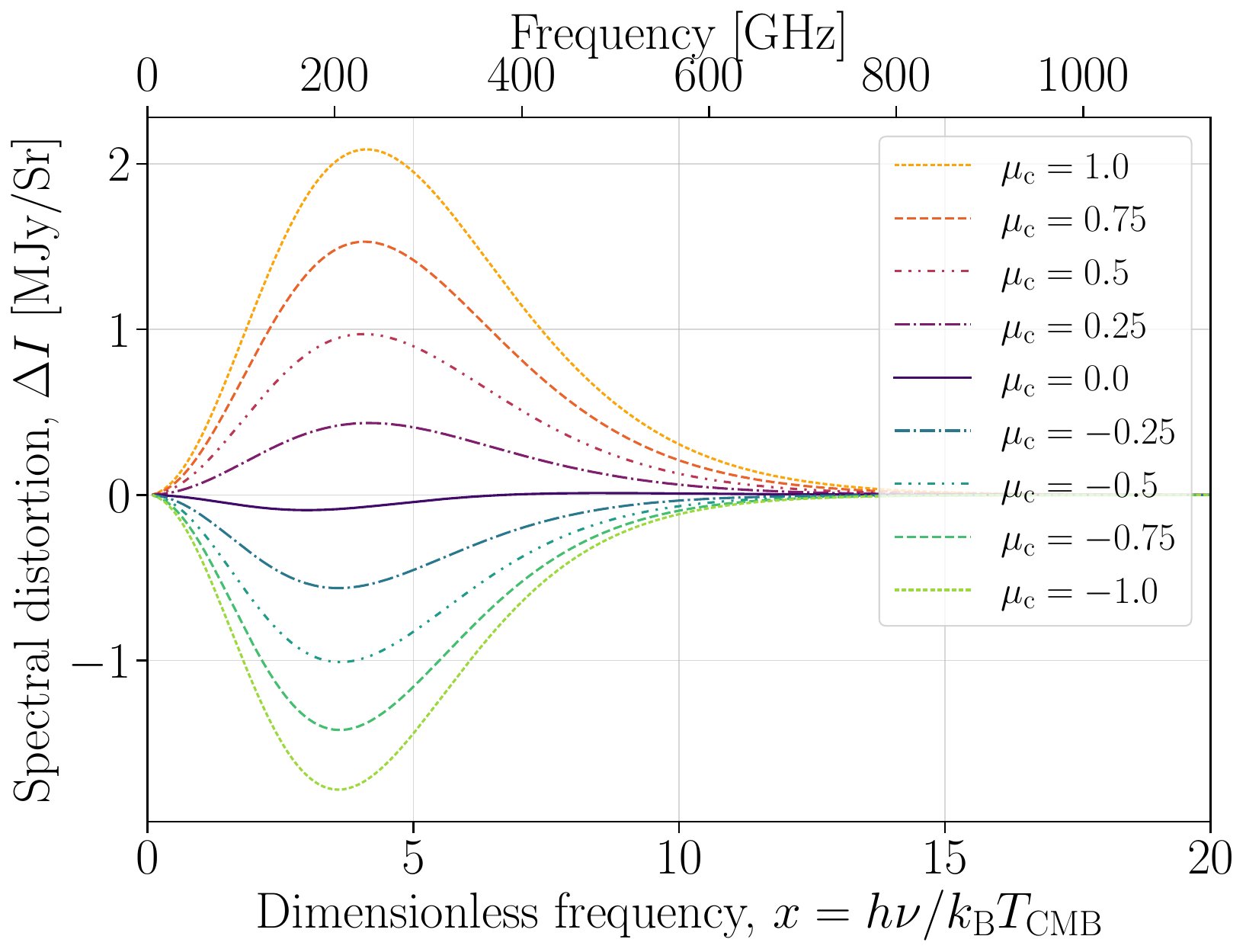}
    \caption[Anisotropic electron beam SZ signals]{The SZ signals induced by a ``beam'' of electrons [to the 5th order in $\ell$, i.e., Eq.~\eqref{eqn:beam_formula}]. The electrons have fixed momentum of $\Bar{p} = 0.1$ and $\muc$ indicates the cosine of the angle between the line-of-sight and the beam. Again we have used $y=10^{-4}$ and $\tau = y/(\betaz^2/3)$.}
    \label{fig:4a_beam_dist_electron_anis}
\end{figure}

We can examine this behaviour in another way, by considering a beam of electrons travelling at a certain angle to the line-of-sight, or alternatively 
%--------------------------------------------------------------------------------
\begin{equation} \label{eqn:beam_formula}
    p^2 f(\pmb{p})=\sum_\ell \frac{2\ell + 1}{4\pi}P_\ell(\muc) P_\ell(\mup)\; \Ne\delta(p-\Bar{p}),
\end{equation}
%--------------------------------------------------------------------------------
with $\Bar{p}$ setting the fixed momentum of the electron beam and $\muc$ the cosine of the angle with respect to the line-of-sight. 
In the expression above, one can identify $p^2 f_{\ell}(p)=\frac{2\ell + 1}{4\pi}\,P_\ell(\muc)\,\Ne\delta(p-\Bar{p})$.
Fig.~\ref{fig:4a_beam_dist_electron_anis}, shows this behavior for the sum up to $\ell\leq 5$, $\Bar{p}=0.1$ and a variety of viewing angles. 

%--------------------------------------------------------------------------------
%--------------------------------------------------------------------------------

For a beam heading directly towards or away from the line-of-sight (that is $\muc=+1$ or $-1$ respectively), the signal is dominated by the dipolar anisotropy. The beam heading directly towards the observer results in a strictly positive signal, increasing the photon count at all frequencies. However, due to the intrinsic anisotropy in the scattering process, a beam heading away is not precisely the opposite, and while it predominantly leads to a decrement in observed photons, at the highest frequencies leads to a very small increment in the photon count for this distribution. 

A beam perpendicular to the line-of-sight (i.e., $\muc=0$) leads to a signal far closer to the conventional SZ signal. This could be expected, as a brief examination of Eq.~\eqref{eqn:beam_formula} shows that only the even $\ell$ contribute in this regime; as such the largest correction comes from the quadrupole ($\ell=2$) component, with a further correction from the $\ell=4$ part. All other angles lie somewhere smoothly between these three angular references, developing more tSZ-like features as the angles approaches $\muc=0$ and stronger kSZ-features at the angular extremes. These illustrations show how general scattering cases with anisotropic electron distributions can be easily obtained using the kernels given above. Averages over the momentum distributions $f_\ell(p)$ can then be carried out numerically.

\subsection{A note on the coordinate systems}
%%%%%%%%%%%%%%%%%%%%%%%%%%%%%%%%%
It is worth noting that in this situation, as all our quantities are being calculated in the frame of an isotropic CMB and an anisotropic electron population, we must consider the physical interpretation of the optical depth and the meaning of the cosine, $\muc$, of the angle between a bulk electron motion and the line-of-sight.

For the latter case it is pertinent only to mention that the angle will experience the relativistic aberration effect, if the rest frame of the electron population is different to that of the CMB. In such a case, if the two frames differ by a velocity $\betac$, then the angle in the electron frame can be written in terms of the angle in the CMB frame as $\mu_{\rm c,e} = (\mu_{\rm c,CMB}-\betac)/(1-\betac\mu_{\rm c, CMB})$.

The effect on the scattering optical depth has been discussed in depth in, e.g., \cite{Chluba2012, Nozawa2013}, and is a frame dependent quantity. For the specific case of the kinematic boost (which will be explored further in the next section) it is simple to identify that an isotropic electron population correlates to a cluster at rest, and as such the optical depth is the optical depth in the cluster frame. However, a cluster travelling at a velocity $\betac$ relative to the CMB, leads to a kinematic correction to the optical depth as measured in CMB frame. In particular, as explained in CNSN one can show that\footnote{Here one uses ${\rm d} t^{\rm c}=\gammac(1-\betac\muc){\rm d} t$ and $\Ne^{\rm c}=\Ne/\gammac$.} 
%----------------
\begin{equation}
\label{eq:tau_trans}
    \tau^{\rm c}_{\rm e}=\int c \sigT \Ne^{\rm c} {\rm d} t^{\rm c}\equiv (1-\betac\muc)\!\int c \sigT \Ne {\rm d} t
    =(1-\betac\muc)\tau_{\rm CMB},
\end{equation}
%----------------
with $\muc$ being the cosine of the angle between the cluster motion and the line-of-sight. This leads to a mixing of multipoles that must be considered if a multipole-by-multipole approach is applied.
If the distortion can be expressed as $\Delta I(x,\muc) = \tau_{\rm CMB} \sum_\ell I_\ell(x) P_\ell(\muc)$, this can also be written as 
%----------------
\begin{equation} \begin{split}
    \Delta I(x,\muc) &= \frac{\tau^{\rm c}_{\rm e}}{(1-\betac\muc)}\sum_\ell I_\ell(x) P_\ell(\muc) 
\end{split} \end{equation}
%----------------
It should, however, be reiterated that it is not always the case that this Lorentz correction should be employed. A careful consideration of the nature of the anisotropy must be carried out, and an understanding of which frame the optical depth is being measured in must be obtained, before determining what, if any, correction must be performed. As argued in CNSN, the easiest and physically most clean way is probably to think of the electron temperature and total scattering optical depth in the cluster rest frame. In a similar way, anisotropies of the electron distribution should be defined with respect to the frame in which the average momentum of the moving electron distribution vanishes. This will allow a clear separation of kinematic corrections to the average optical depth from distribution function anisotropies and their parametrization. We believe this issue is also at the core of the discussion in \cite{Molnar2020}, which address the question about the correct J\"uttner distribution \citep[e.g.,][for introduction to the topic]{Dunkel2007}.

%--------------------------------------------------------------------------------
\subsection{Kinematic boost using anisotropic electron kernels}
\label{sec:kSZ_electrons}
In most situations, we would not expect the jet-like electron distribution to be monoenergetic. As such, it is worth briefly discussing the kinematic effects as derived through this formalism. In the CMB rest frame, a cluster moving at a speed $\betac$ with direction cosine $\muc$ and the intrinsic\footnote{This refers to the temperature as measured in the cluster frame, although we do not explicitly distinguish this temperature. For some discussion on definitions of the temperature see Appendix~C1 of \cite{Chluba2012}.} temperature $\thetae=\kB \Te/\me c^2$ can be expressed as a boosted thermal distribution. In particular, a regular relativistic Maxwell-Boltzmann distribution\footnote{That is, a Maxwell-J\"uttner distribution.} can be written as
%------------------------------
\begin{equation} \label{eqn:rMB_dist}
    f^{\rm c}_{\rm th}(\pmb{p}^{\rm c}) =\frac{N^{\rm c}_{\rm e} \expf{-\gamma^{\rm c}/\theta_{\rm_e}}}{4\pi (m_{\rm e} c)^3 K_2(1/\thetae) \thetae}
\end{equation}
%------------------------------
with $\gamma^{\rm c}$ in the restframe of the cluster. 
This distribution is normalized as $N^{\rm c}_{\rm e}=\int {\rm d}^3 p^{\rm c}\,f^{\rm c}_{\rm th}(\pmb{p}^{\rm c})$.
To obtain the required expression in the CMB frame, we can use that the electron number ${\rm d} N^{\rm c}_{\rm e}=f^{\rm c}_{\rm th}\,{\rm d}^3 x^{\rm c} \,{\rm d}^3 p^{\rm c}$ is invariant. Since ${\rm d}^3 x^{\rm c} \,{\rm d}^3 p^{\rm c}$ is invariant, also $f^{\rm c}_{\rm th}$ has to be invariant. One therefore only has to express the electron energy in terms of variables in the CMB frame.
%--------------------------------------------------------------------------------
\begin{figure}
    \centering
    \includegraphics[width=0.8\linewidth]{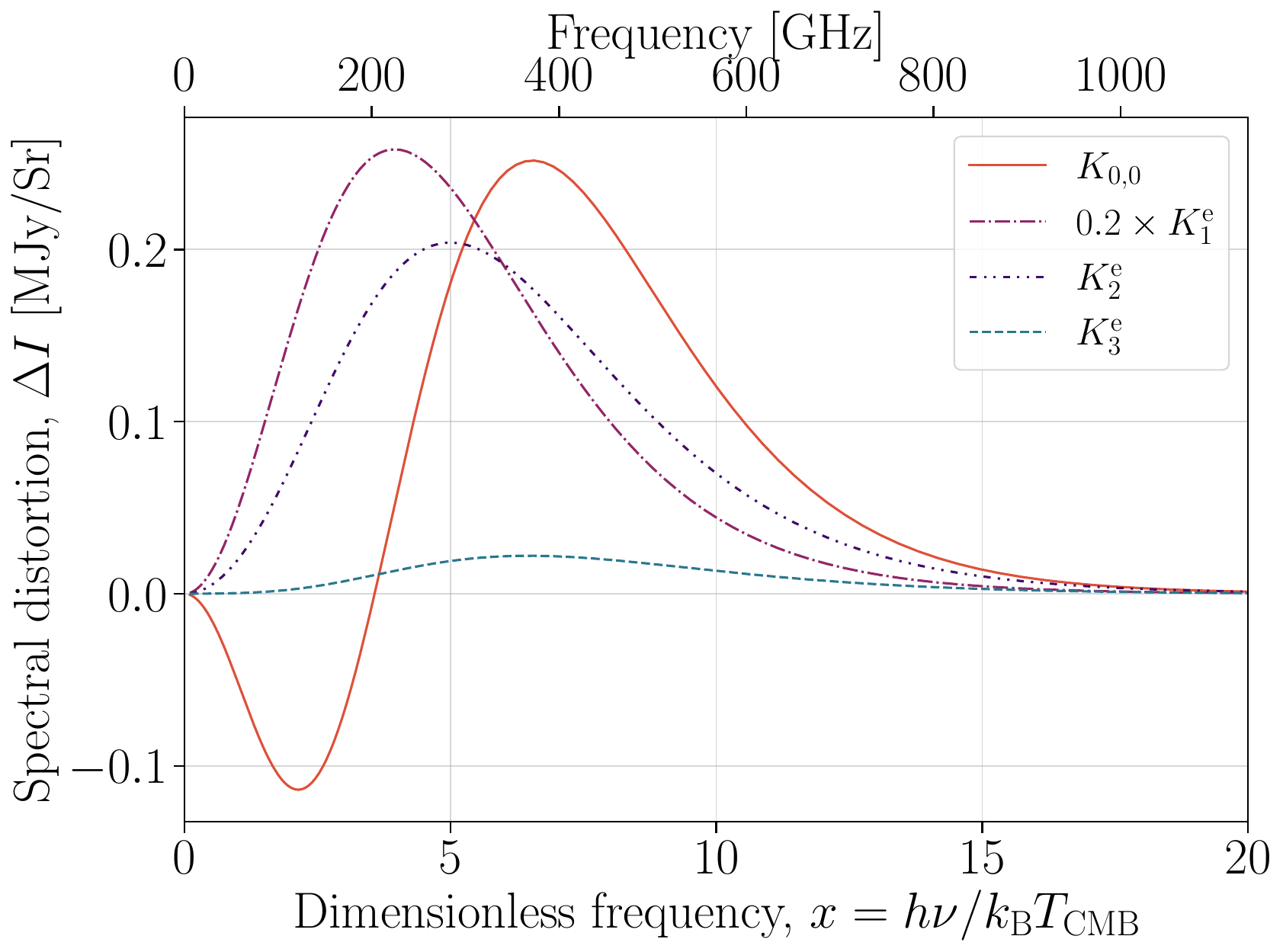}
    \caption[Kinematic electron SZ signals]{The kinematic SZ signals for each anisotropy -- that is the signal induced by a boosted thermal distribution. Here, a 5~keV thermal electron distribution is used, boosted with a kinematic correction of $\betac=0.1$, and $\muc=1$. Again we have used $y=10^{-4}$, except the dipole term which is suppressed with a factor of $0.2$.}
    \label{fig:5_boost_dist_electron_anis}
\end{figure}
%--------------------------------------------------------------------------------

It is easy to show \citep[e.g.,][]{Nozawa1998SZ}, that the directional boost can be expressed through\footnote{This can be shown using the velocity addition theorem in special relativity.} $\gamma^{\rm c} = \gammac(\gamma-\betac p\,\mu_{\rm e, c})$ where $\gamma$ and $p$ are evaluated in the CMB frame, and $\betac$ the relative speed\footnote{Hence, $\gammac=1/\sqrt{1-\betac^2}$ is the associated Lorentz factor.} of the the cluster and CMB frames. In addition, $\mu_{\rm e, c}$ here is the cosine of angle between the incoming electron and the axis of bulk motion. Then the boosted distribution function can be expressed as 
%------------------------------
\begin{equation} \label{eqn:rMB_dist_boost}
    f_{\rm boost}(p) =
  f^{\rm c}_{\rm th}(\gammac\gamma)\times \expf{\frac{\mu_{\rm e, c}\pc p}{\thetae}}
  =
\frac{(N_{\rm e}/\gammac)\, \expf{-\gammac\gamma/\theta_{\rm_e}}}{4\pi (m_{\rm e} c)^3 K_2(1/\thetae) \thetae}
\times \expf{\frac{\mu_{\rm e, c}\pc p}{\thetae}}
\end{equation}
%------------------------------
with $N^{\rm c}_{\rm e}=N_{\rm e}/\gammac$, which ensures that relation~\eqref{eq:tau_trans} directly applies without further ado.\footnote{Without this factor of $1/\gammac$ one would find $\tau_{\rm e}=\gammac(1-\betac\muc)\tau_{\rm CMB}$, which is not the correct transformation law (see CNSN). This factor can also be understood from the fact that $N^{\rm c}_{\rm e}$ has to transform as the inverse of volume, confirming $N^{\rm c}_{\rm e}=N_{\rm e}/\gammac$ \citep[e.g.,][]{Sazonov1998}.} 
We note in passing that carrying out the integral of $f_{\rm boost}(p)$ over ${\rm d}^3 p$ will yield $N_{\rm e}$.
The second exponential term can then be expanded in Legendre polynomials as
%--------------------------------------------------------------------------------
\begin{equation} \label{eqn:boostMB}
    \expf{A \mu_{\rm e, c}} = \sqrt{\frac{\pi}{2A}}\sum_{\ell=0}^\infty (2\ell+1) I_{\ell+\frac{1}{2}}(A) P_\ell(\mu_{\rm e, c}),
\end{equation}
%--------------------------------------------------------------------------------
with $A=\pc p/\thetae$ and where $I_\ell(x)$ is the modified Bessel function of the first kind -- which is a modification of the expression found in \cite{Hu1997}. 
This naturally leads to an anisotropic electron distribution due to the bulk motion. To compute the scattered signal, we can use spherical harmonic addition theorem to express  $P_\ell(\mu_{\rm e, c})$ with $\mu_{\rm e, c}=\muc\mu'+\cos(\phi_{\rm c}-\phi') (1-\muc^2)^{1/2}(1-{\mu'}^2)^{1/2}$. It is worth highlighting, that then, in the case of an isotropic photon distribution, it is equivalent to use the conversion $P_\ell(\mu_{\rm e, c})\rightarrow P_\ell(\muc)P_\ell(\mu')$.
When substituting in this electron distribution for each anisotropy, we can recreate the kinematic SZ corrections by using the anisotropic electron scattering kernels and multiplying the result by $1/(1-\betac \muc)$. This is because in the CMB frame the line of sight integral will yield a factor of $\tau_{\rm CMB}=\tau^{\rm c}_{\rm e}/(1-\betac \muc)$ according to Eq.~\eqref{eq:tau_trans}.

%--------------------------------------------------------------------------------
\begin{figure}[H]
    \centering
    \includegraphics[width=0.7\columnwidth]{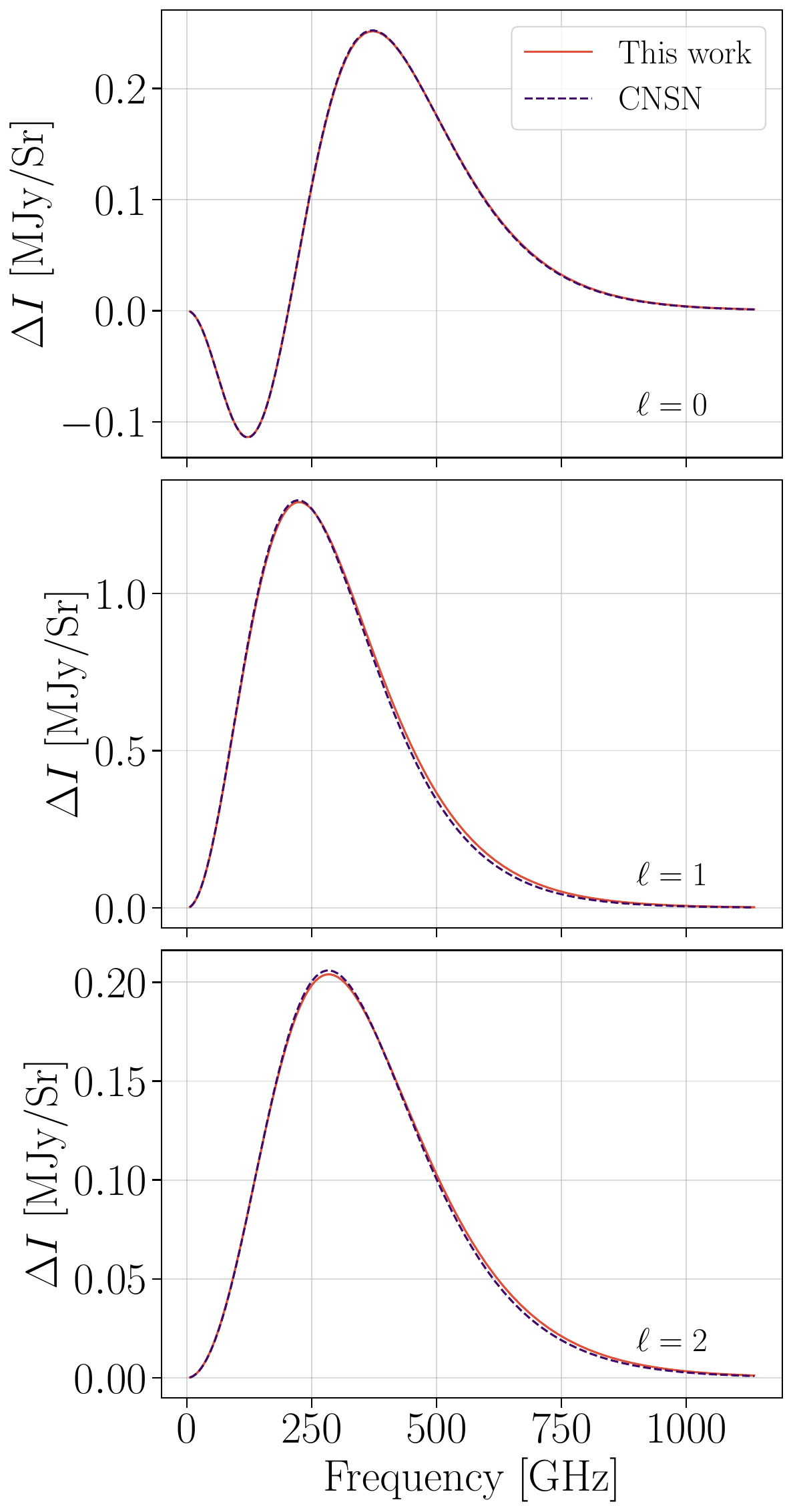}
    \caption[Kinematic electron vs CNSN SZ signals]{
A comparison of signals on a multipole basis generated by the boosted electron distribution ``This work'' and the transformed CMB signal described in CNSN. Here $\betac=0.1$, $\muc=1.0$ and $\ell$ refers to which multipole is generating the signal. That is, the monopole, dipole and quadrupole signal from top to bottom respectively.}
    \label{fig:5b_boost_vs CNSN}
\end{figure}
%--------------------------------------------------------------------------------

In principle, the signals obtained using this procedure should be identical to the kinematic corrections described in Section~\ref{sec:kSZ_photons}, 
and indeed we can see this is largely correct in Fig.~\ref{fig:5_boost_dist_electron_anis}, where the first four multipoles are displayed for $\betac=0.1$ and $\muc=+1$. This is further validated in Appendix~\ref{app:kSZ_comp}, where the combination of these multipole components are compared to a full kinematic SZ signal. It should be noted that, while this may be a reasonable velocity for high-resolution considerations of the SZ effect -- where small volume elements may be boosted by jets or shocks -- for full cluster considerations of kinematic effects, far lower energies are generally expected. In particular, typical cluster velocities are $\mathcal{O}(100)$~km~s$^{-1}$, that is $\betac \approx\mathcal{O}(10^{-3})$. Nevertheless, our calculations demonstrate the validity of the electron kernel approach and highlight that the method can be reliably used to consider extreme scenarios.

We can further compare the results visually to the forms of the kinematic effects provided in e.g., CNSN. This is displayed in Fig.~\ref{fig:5b_boost_vs CNSN}. We can immediately see that to first order these are very closely aligned despite the high $\betac$ value being considered. The small differences identifiable in the dipole and quadrupole components can be understood as an effect of the curtailing of the CNSN method to $\mathcal{O}(\betac^2)$ while the expressions in Eq.~\eqref{eqn:boostMB} are complete in all orders of $\betac$. In Appendix~\ref{app:kSZ_betacexp} we explore the effect of $\betac$ order, to show that this is the entire cause of the discrepancy seen here.

The dipole and quadrupole, as such, exhibit similar shapes to those presented CNSN, although ever so slightly broader at high $\betac$. The octopole term displayed only in Fig.~\ref{fig:5_boost_dist_electron_anis} [and neglected in CNSN as it is leading order $\mathcal{O}(\betac^3)$] is a similar shape again, with the peaks of the distributions of the dipole, octopole and quadrupole shifting to higher frequencies with higher $\ell$. It is also worth highlighting again, that the magnitude of the signal generated by the electron anisotropies decrease with increasing $\ell$ -- with the exception of the dipole term compared to the monopole, where a high energy boost may be larger than the tSZ signal contained in the monopole, where the first $\betac$ terms come at $\mathcal{O}(\betac^2)$ [the monopole corrections discussed in CNSN].

It should also be mentioned here that for a simple kinematic boost, the approximation $f(\ppb)\approx f(p_0, \ppbh)$ used in Eq.~\eqref{eq:def_gen} can be explicitly confirmed. Since for the considered case we have 
%--------------------------------------------------------------------------------
\begin{align} 
\label{eqn:boostMB_relation}
\frac{f(\ppb)}{f(\pzb)}&=
\exp\left(\frac{\gammac(\gamma_0-\gamma')+\pmb{p}_{\rm c}\cdot(\pmb{p}'-\pmb{p}_0)}{\thetae}\right)
\nonumber\\
&=\exp\left(\frac{\gammac T_{\rm CMB}(x'-x_0)}{T_{\rm e}}+\frac{\pmb{p}_{\rm c}\cdot(\pmb{k}_0-\pmb{k}')}{\thetae}\right),
\end{align} 
%--------------------------------------------------------------------------------
we can safely drop the terms $\propto T_{\rm CMB}/T_{\rm e}\simeq 2.4\times 10^{-4} \,{\rm eV}/5 \,{\rm keV} \simeq 5 \times 10^{-8}$. Similarly, we have $(\pmb{k}_0-\pmb{k}')/\thetae\simeq (x_0-x') T_{\rm CMB}/T_{\rm e}$, leaving a negligible correction. For similar reasons we were able to neglect stimulated scattering terms which scale as $\simeq n(1+n)\, T_{\rm CMB}/T_{\rm e}$. These approximations greatly simplify the otherwise more complex coupling to the electron and photon anisotropies.

%--------------------------------------------------------------------------------
\section{High Energy electron populations}
\label{sec:hEnT}
%--------------------------------------------------------------------------------
In the previous sections, we primarily demonstrated the utility of the anisotropic scattering kernel approaches for cases closely related to the thermal and kinematic SZ effects.
It is now worthwhile to examine high-energy {\it non-thermal} electron distributions and their interactions with the monopole and anisotropic scattering effects, to determine if these effects are meaningfully magnified or modified in these regimes.
In particular, we examine five electron distribution models that have been considered in the literature. The first is the standard relativistic Maxwell-J\"uttner distribution used generally in rSZ described in Eq.~\eqref{eqn:rMB_dist}.

%--------------------------------------------------------------------------------
\begin{figure}[H]
    \centering
    \includegraphics[width=0.9\columnwidth]{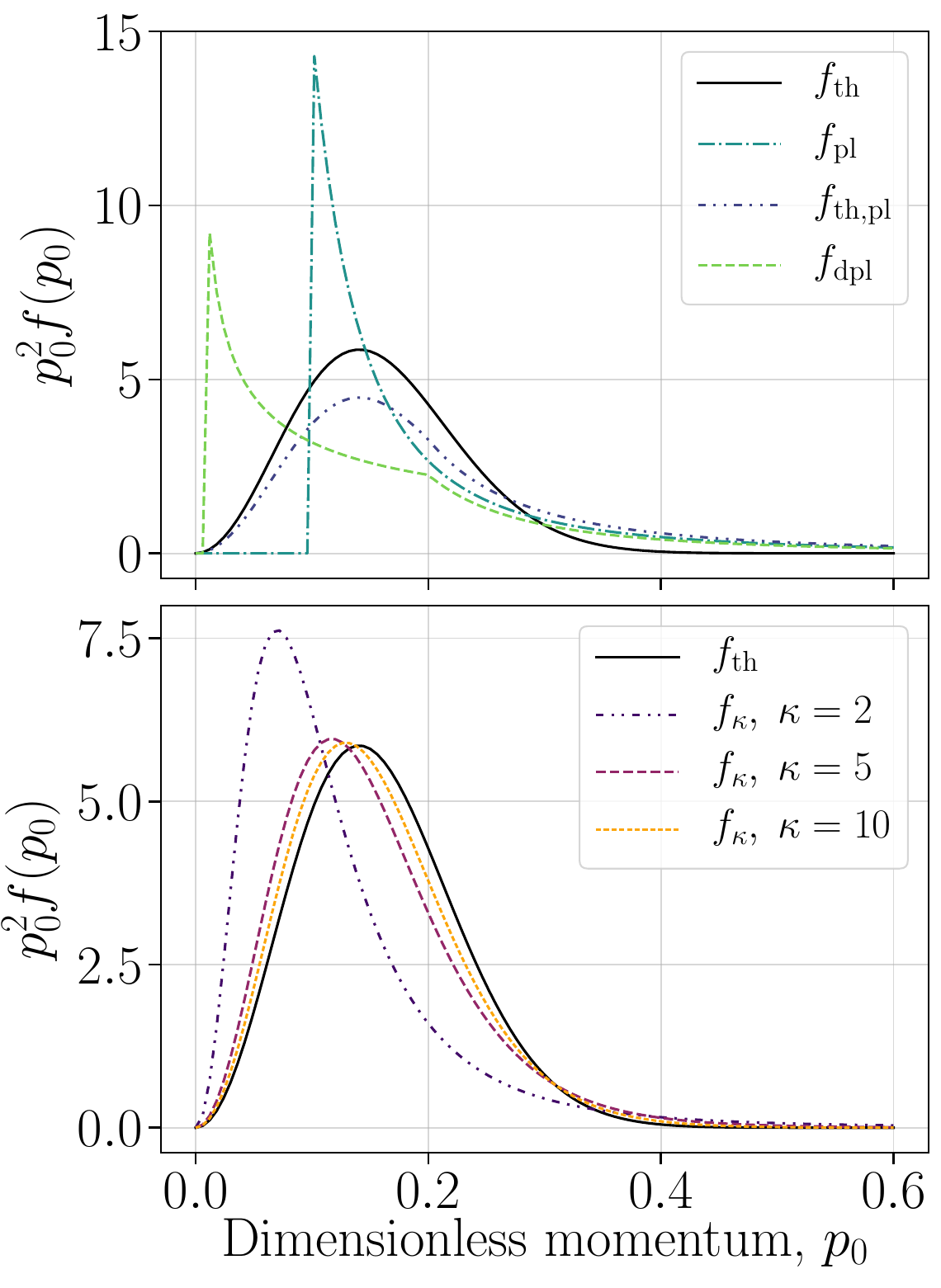}
    \caption[Toy models for electron distributions]{A selection of toy models for the high energy electron momentum distributions found in clusters. In black, in both figures is the thermal, relativistic Maxwell-Boltzmann distribution at $5$~keV. This temperature is used for all the other temperature dependent models (except where indicated otherwise). {\it Top}: The power law distributions, here in particular, $f_{\rm pl}(\pz;2.5,0.1,10)$ (Eq.~\ref{eqn:f_pl}), $f_{\rm th,pl}(\pz;5,2.5,0.2,10)$ (Eq.~\ref{eqn:f_thpl}) and $f_{\rm dpl}(\pz;0.5,2.5,0.01,0.2,10)$ (Eq.~\ref{eqn:f_dpl}). {\it Bottom}: The $\kappa$-distributions $f_\kappa(\pz; 5, 2)$ , $f_\kappa(\pz; 5, 5)$ and finally $f_\kappa(\pz; 5, 10)$ (Eq.~\ref{eqn:f_kappa}).}
    \label{fig:6_electron_funcs}
\end{figure}
%--------------------------------------------------------------------------------

The next is a simple power law, as used in \cite{Ensslin2000},
%--------------------------------------------------------------------------------
\begin{equation} \label{eqn:f_pl}
    p^2 f_{\rm pl}(p;\alpha,p_1,p_2) = \frac{(\alpha-1)p^{-\alpha}}{p_1^{1-\alpha}-p_2^{1-\alpha}} \quad {\rm for \;} p_1<p<p_2.
\end{equation}
%--------------------------------------------------------------------------------
In that work, they also introduced a hybrid distribution that uses the thermal distribution at low frequencies and a power law at higher frequencies,
%--------------------------------------------------------------------------------
\begin{equation}\label{eqn:f_thpl}
    p^2 f_{\rm th,pl}(p;\theta, \alpha,p_1,p_2) \propto
    \begin{cases}
        f_{\rm th}(p,\theta), & {\rm if\; } p\leq p_1\\
        f_{\rm th}(p_1,\theta)\left(p/p_1\right)^{-\alpha},& {\rm if\; } p_1<p<p_2
    \end{cases}
\end{equation}
%--------------------------------------------------------------------------------
where a normalisation function $C(\theta, \alpha,p_1,p_2)$ is defined such that $\int p^2 f_{\rm th,pl} {\rm d}p = 1$. 

In \cite{Colafrancesco2002}, the consideration of power laws was extended to a double power law model, to reduce the contribution at the lower energies, where a single power law became distinctly unphysical. That is, 
%--------------------------------------------------------------------------------
\begin{equation}\label{eqn:f_dpl}
    p^2 f_{\rm dpl}(p;\alpha_1,\alpha_2,p_1,p_2, p_{\rm cr}) = 
    \begin{cases}
        p^{-\alpha_1}, & {\rm if\; } p_1< p \leq p_{\rm cr}\\
        p_{\rm cr}^{-\alpha_1+\alpha_2} p^{-\alpha_2},& {\rm if\; } p_{\rm cr}<p<p_2,
    \end{cases}
\end{equation}
%--------------------------------------------------------------------------------
where again a normalisation function $K(\alpha_1,\alpha_2,p_1,p_2, p_{\rm cr})$ is defined to normalise the distribution. 

Moving away from power laws, \cite{Kaastra2009} uses two generalisations of the maxwellian model, of which we will use the $\kappa$-distribution, which tends to a maxwellian as $\kappa\rightarrow\infty$,
%--------------------------------------------------------------------------------
\begin{equation}\label{eqn:f_kappa}
    p^2 f_{\kappa}(p;\theta,\kappa) = \frac{\frac{\sqrt{2}p^2}{\sqrt{\pi}\theta^{3/2}}\frac{\Gamma(\kappa+1)}{(\kappa-1.5)^{3/2}\Gamma(\kappa-0.5)}}{\left(1+\frac{p^2}{2\theta(\kappa-1.5)}\right)^{\kappa+1}},
\end{equation}
%--------------------------------------------------------------------------------
with $\kappa$ an integer and $\Gamma$ the gamma functions.

These models can all be seen graphically in Fig.~\ref{fig:6_electron_funcs}. Here we use a variety of small values of $\kappa$ for the $\kappa$-distribution. We have only plotted illustrative values of the power law related distributions.

%--------------------------------------------------------------------------------
\subsection{Illustrations of the Monopole signal}
\label{sec:nt_monopole}

\begin{figure}[H]
    \centering
    \includegraphics[width=0.9\columnwidth]{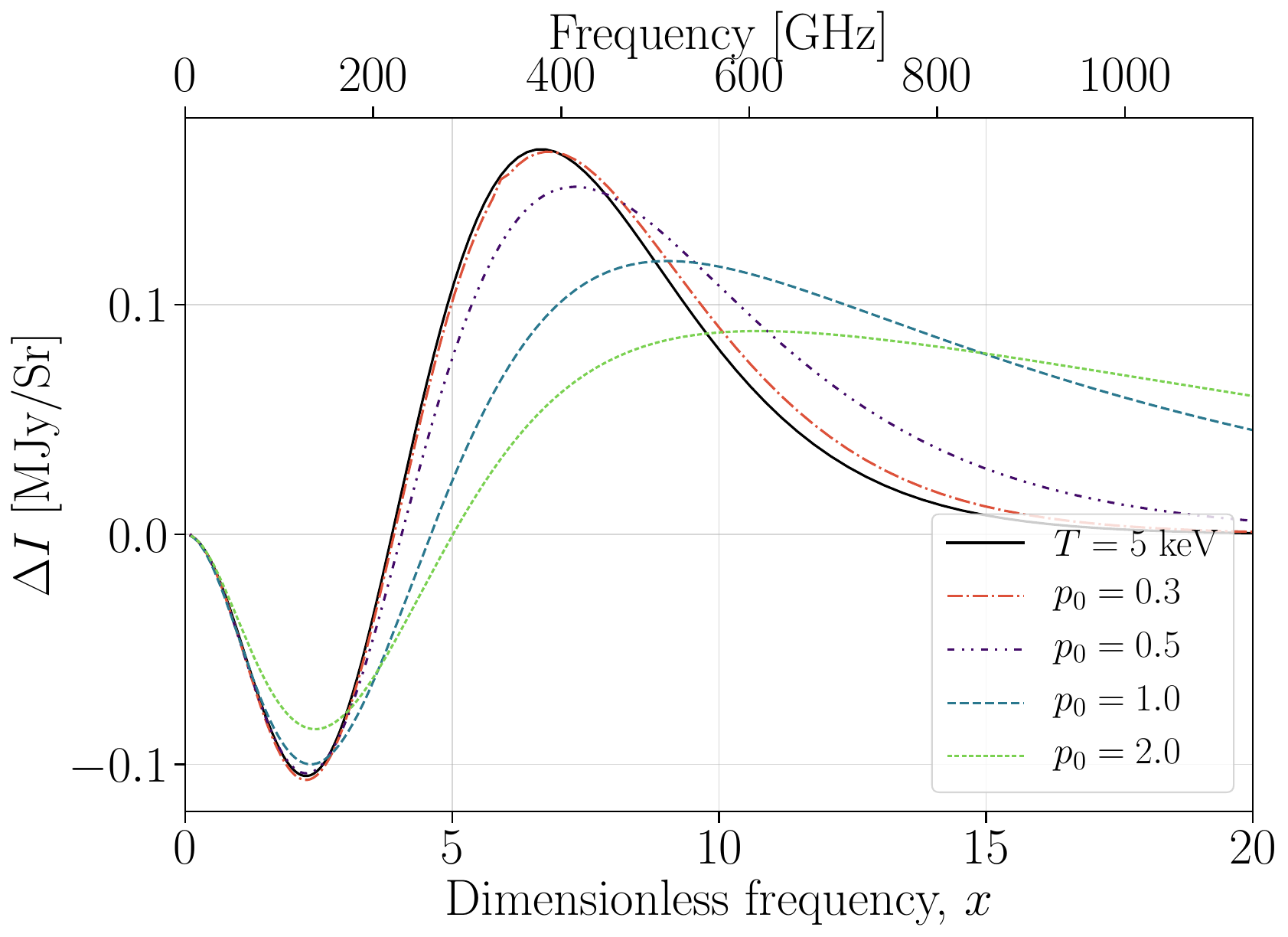}
    \caption[Spectral distortion from fixed momenta electrons]{The distortion shapes for a selection of fixed momentum values. Here $y=10^{-4}$, the thermal distribution in black is at $5$~keV, and for the fixed momenta, the optical depth is calculated as $\tau=y/(\betaz^2/3)$.}
    \label{fig:7a_dists_fixed momenta}
\end{figure}

To understand the effects of high energy contributions to the SZ effect, it is first instructive to consider the effects of fixed momenta regions. This is generally unphysical, but allows for insights into the effects of non-thermal components. In particular, in Fig.~\ref{fig:7a_dists_fixed momenta}, we have plotted the distortion effect of a variety of electron momenta. At higher momenta, the distributions become distinctly non-thermal, with extended and increased tails to higher frequencies. These momenta correspond to relativistic speeds that boosted electrons may reach in clusters, with $\pz=1$ ($\gammaz\simeq 1.4$) equating to electrons travelling at half the speed of light.

%---------------------------------------------------------------------------------
%---------------------------------------------------------------------------------

\begin{figure}[H]
    \centering
    \includegraphics[width=0.8\columnwidth]{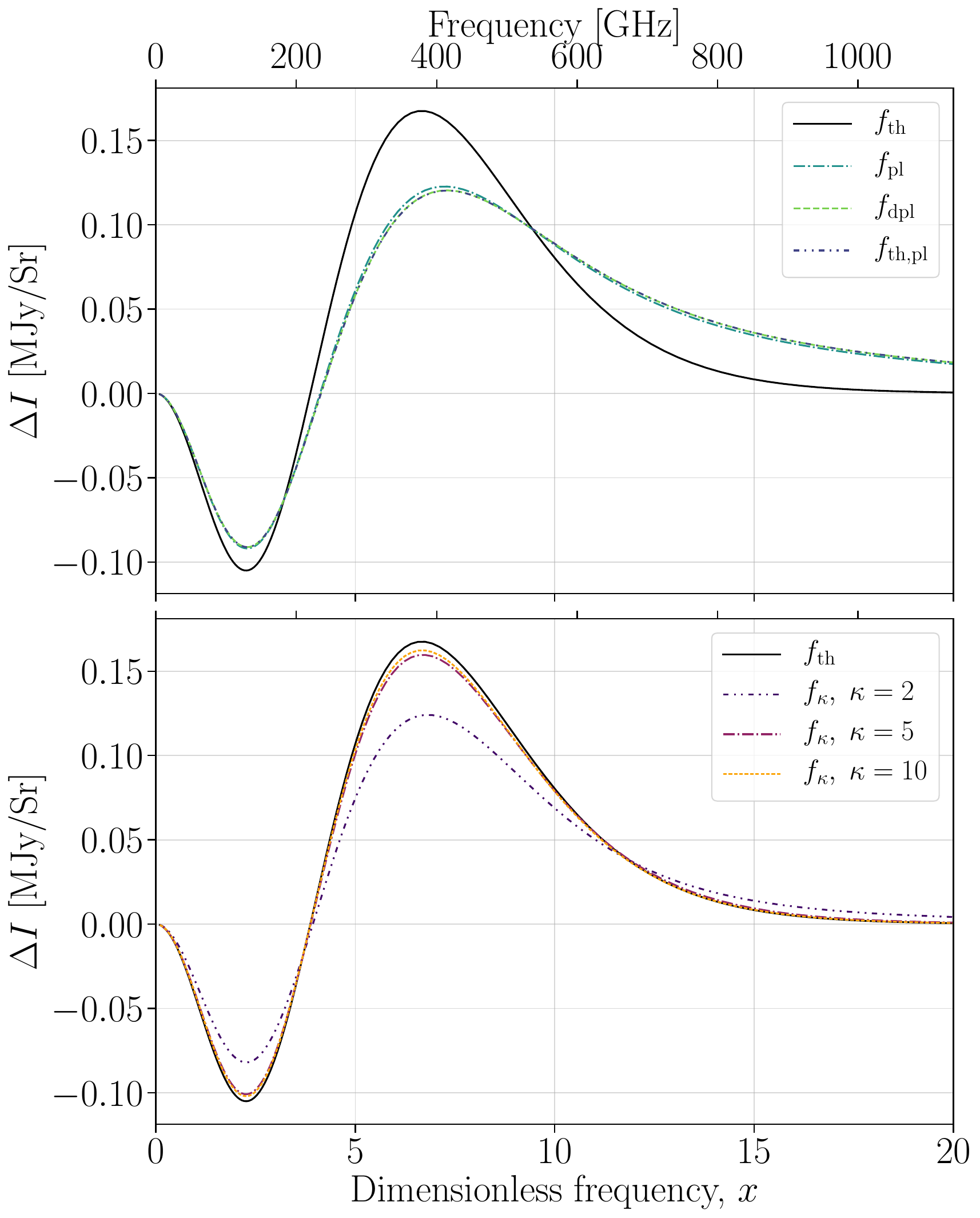}
    \caption[Spectral distortion from different electron distributions]{The distortion shapes for the distributions displayed in Fig.~\ref{fig:6_electron_funcs}. Everywhere, where a temperature is required in the distribution, 5~keV is used and $y=10^{-4}$ is also assumed. The functions including power laws {(in the upper panel)} all have $\tau = y/(\overline{\beta^2}/3)$, where $\overline{\beta^2}$ is the averaged dimensionless velocity squared within the distribution.}
    \label{fig:7_dists_from_electrons}
\end{figure}

Now, we can consider how the various model distributions lead to variations in the observed SZ signal. Considering the scattering of the CMB monopole, these are all displayed in Fig.~\ref{fig:7_dists_from_electrons}.
It is first worth making a note regarding the amplitudes of these signals -- that is, how we have fixed the optical depth. For the power law signals, we set $\tau = 10^{-4}/\langle\beta^2/3\rangle$ where $\langle\beta^2/3\rangle = \int (\beta^2/3)p^2 f(p) \D{p}$. That is, we calculate the effective temperature\footnote{Note that the effective temperature for $f_{\rm pl}$, $f_{\rm dpl}$ and $f_{\rm th,pl}$, are $\theta_{\rm eff} = 0.0248$, 0.0200 and 0.0290 respectively.} of the distribution and set $\tau$ so that the Compton-$y$ parameter = $10^{-4}$. For the thermal distributions and $\kappa$-distributions, as they explicitly have a temperature dependence, we instead use $y=10^{-4}=\tau\times\theta$.

Now we can notice that the effects of the \cite{Kaastra2009} models, i.e., those in the lower panel, largely have the effect of modelling the relativistic corrections to the SZ effect -- that is, since these models start from the non-relativistic Maxwell-Boltzmann distributions, the effects of these toy-models are broadly comparable to those derived from the relativistic Maxwell-J\"uttner distribution.

The power law models, however, result in a far more distinct signal, although variations between the models are not as significant. In particular, these again give rise to substantially increased tails to the non-thermal SZ distortion. It should be noted that at the lower frequencies, these are very small effects, keeping the negative parts (i.e., $<217$~GHz) almost unchanged. They result in a shift to the null, comparable to high rSZ corrections, and then large changes in the tail, where the amplitude has been greatly increased.

%--------------------------------------------------------------------------------
%--------------------------------------------------------------------------------

%--------------------------------------------------------------------------------
\begin{figure}[H]
    \centering
    \includegraphics[width=0.8\columnwidth]{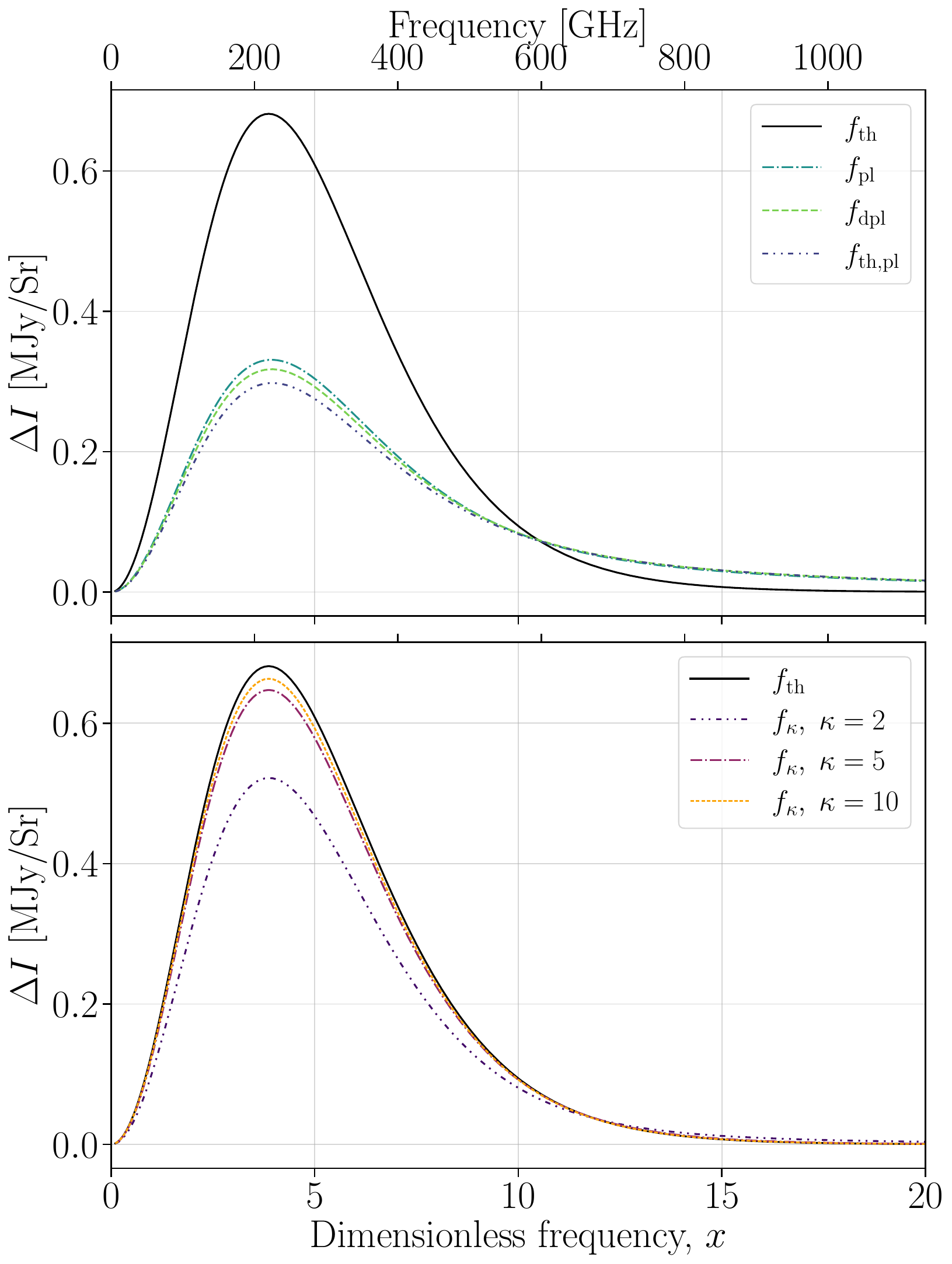}
    \caption[Spectral distortion from different dipole anisotropic electron distributions]{The distortion shape of a purely dipole electron distribution, with the momenta distributions displayed in Fig.~\ref{fig:6_electron_funcs}. The figure is arranged as in Fig.~\ref{fig:7_dists_from_electrons}.}
    \label{fig:8_dipole_nonThermal}
\end{figure}
%--------------------------------------------------------------------------------

\subsection{High-energy non-thermal anisotropies}

\begin{figure}[H]
    \centering
    \includegraphics[width=0.8\columnwidth]{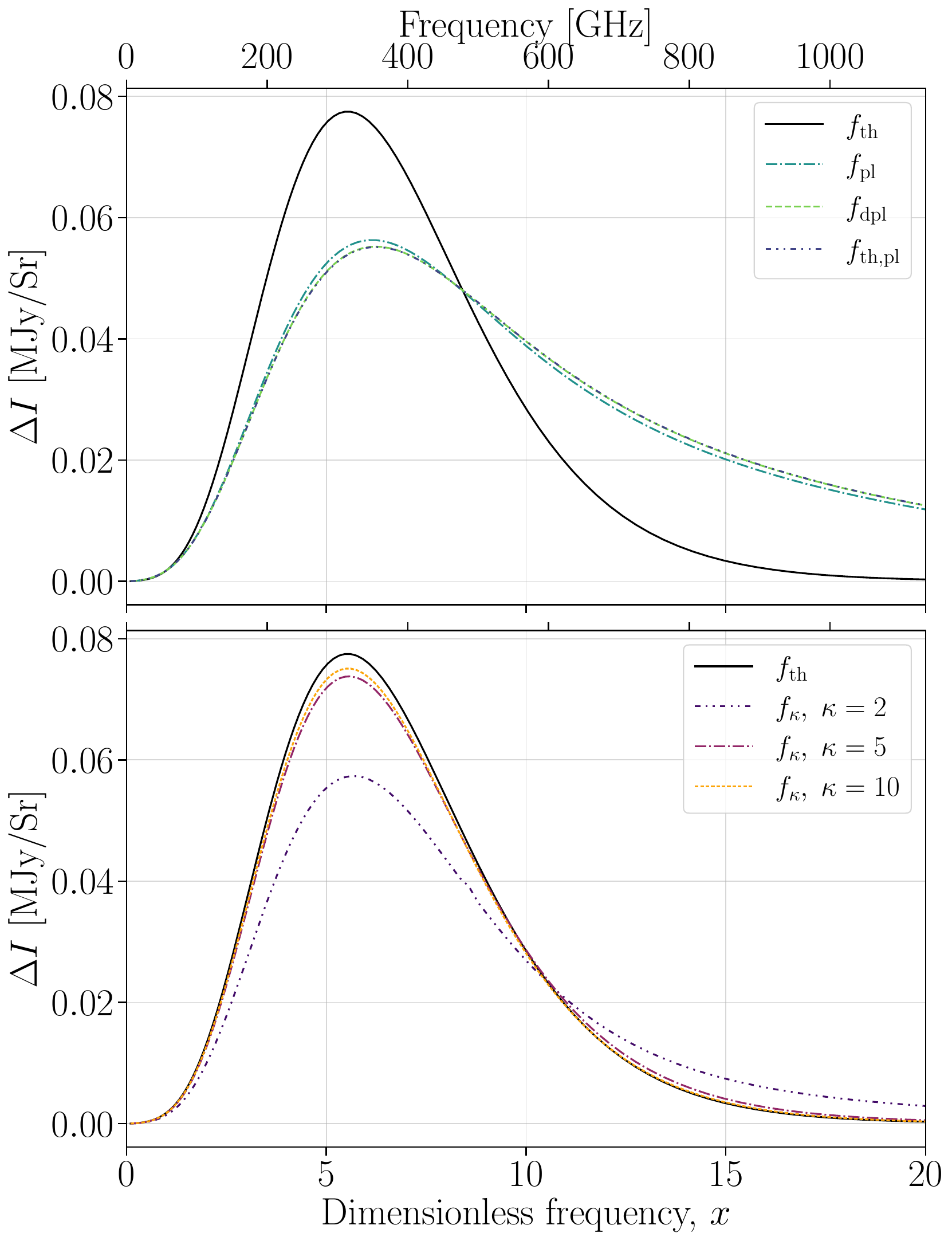}
    \caption[Spectral distortion from different quadrupolar anisotropic electron distributions]{The distortion shape of a purely quadrupolar electron distribution, with the momenta distributions displayed in Fig.~\ref{fig:6_electron_funcs}. The figure is arranged as in Fig.~\ref{fig:7_dists_from_electrons}.}
    \label{fig:8_quadrupole_nonThermal}
\end{figure}

It is natural to extend this consideration to anisotropic electron contributions. For illustration, here we shall consider the same momentum distributions as for the monopole signal just described -- that is, a dipole contribution of the power law distribution would represent a locally anisotropic distinctly non-thermal electron population. The corresponding SZ signals caused by electron dipole and quadrupole contributions are displayed in Figures~\ref{fig:8_dipole_nonThermal} and \ref{fig:8_quadrupole_nonThermal} respectively.

The effects are comparable to those observed in the monopole case already discussed -- the power laws provide smaller peaks with significantly longer tails to high frequencies, while the $\kappa$ distributions show increasing convergence with the thermal distribution as $\kappa$ increases. 

%--------------------------------------------------------------------------------
%--------------------------------------------------------------------------------

In particular, we see that the dipole contribution is greatly suppressed at lower frequencies for these power-law distributions leading to a peak around half the height of that seen for the thermal distribution, with strong variation between each distribution. However, at higher frequencies $\gtrsim600$~GHz, the two signals become comparable and the power-law distributions provide a far stronger tail at higher frequencies.
The $\kappa$ distributions show a similar trend where the more divergent electron distributions lead to more diminished peaks in the scattering distribution and stronger tails to high frequencies.

For a quadrupolar electron anisotropy, the $\kappa$ distributions show a similar tendency again, albeit with starker changes than those seen in the dipole distribution. As $\kappa$ increases, the peak of the scattered distribution falls and the strength of the high frequency tail (once more $\gtrsim600$~GHz) increases.
This can be seen even more strongly in the power-law distributions, where the quadrupole obtains an exceedingly pronounced skew to its distributions. The intensity peak drops and shifts to a slightly higher frequency, and the tail is greatly magnified from around $\sim 500$~GHz.

These variations can be understood, to an extent, through a consideration of the underlying distributions. In particular, the high frequency part of each distortion is driven by the high momenta tail of the power-law distributions where the distributions dominate over the thermal distribution. At lower frequencies the similarities between each signal and the decrement compared to the thermal signal are less well understood.
It should be noted however, that these power-law distributions all use an effective temperature to calculate $\tau$ as described in Section~\ref{sec:nt_monopole}. As such, these comparisons of amplitude may be limited compared to the results derived through a detailed study, and more attention should be paid to the differences in the spectral responses.

As for the $\kappa$ distributions, we see, as would be expected that the high $\kappa$ lead to only small variations from the thermal distribution, matching the similarities between the underlying momenta distributions. However, for $\kappa=2$ where the momentum distribution has a significantly sharper peak at lower momenta, we can see larger differences in the resulting distortion. In general, for both the dipole and quadrupole signals we see a decrease in signal at lower frequencies (for the same optical depth), but the high energy tail is largely unaffected. This lower frequency variation does not lead to a large variation in the shape, but leaves the peak troughs and null-crossings at almost the same frequencies. Instead it merely reflects a weaker scattering for the same inputted $\tau$.

While these effects are all only modelled through toy-model predictions for non-thermal behaviour in clusters, some conclusions can still be drawn. Firstly, when calculating these non-thermal effects, increased resolution and precision in measurement of clusters must be obtained than is typically available. However, these distortions are generally comparable to the rSZ corrections, either in significance of the effect, or in that they are dominant at the higher frequencies that are less accessible, likely requiring space-based observation. 

That said, more study must be done into the proportion of of electrons residing at these high energies, and the electron distributions they are residing in. For instance, if the high energy non-thermal electrons account for 1\% of the ICM, these signals would result in distortions around 1\% of the scale of the rSZ corrections.
However if these electrons contribute significantly more, or have extremely relativistic distributions, this may result in a larger distortion which may be sufficient to be observed.

Furthermore, for higher resolution studies, where regions of high energy electrons can be distinguished more clearly (i.e., jets, shocks, etc.) then the ICM in these regions will likely have a larger non-thermal component, and thus an increased distortion impact.
We present here an illustrative method for directly calculating these signals for arbitrary underlying distributions, allowing for future studies to directly input non-thermal anisotropic momentum distributions to determine the specific shapes generated and evaluate the variation from a purely classical tSZ signal.

%--------------------------------------------------------------------------------
\section{Discussion}
\label{sec:discussion}
%--------------------------------------------------------------------------------
\subsection{Observational capacity}
It is difficult to use these calculations to determine the scale of the effects from non-thermal and anisotropic distributions without stronger and more realistic models for both the underlying electron and photon distributions and estimates for their prevalence within clusters.
Indeed, the main motivation for our work here is to provide the required tools to model the related SZ signals efficiently.
However, there are still several conclusions that can be drawn. In particular, these effects will always be more prevalent in high resolution and high precision measurements of clusters, as, in general, we are discussing localised sources of non-thermality and anisotropy. Where sources lead to large-scale changes to the electron distributions over the whole cluster, signals will be more visible and more relevant for the accurate determination of fundamental constants of the clusters themselves.

Secondly, it is worth noting that different non-thermal models can modify either the low or high energy portion of the underlying signals, meaning that different distributions will be measurable to different extents depending on the experiment/observatory being used to measure potential SZ signals. Further work must therefore be carried out to determine which kinds of non-thermal models are the most appropriate for each source of non-thermal electrons in clusters, to determine how each will affect the observational potential.

Additionally, very little work has been carried out to determine the levels of anisotropy within electron distributions (beyond kinematic bulk motion corrections). As such these should be the focus of further study which might indicate improved estimates of the expected scale of their contribution. These will rely on continued modelling of the magnetic fields in clusters and determining their influence on the microphysical anisotropy of the electrons in the ICM. Furthermore, high resolution studies may increase the detectability of these effects allowing for closer study of the regions of strong anisotropy, and contribute to our understanding of the behaviour of magnetic fields in clusters.
Moreover, since the dipole components can have large signals that are comparable in size to monopole signal, it is very important to consider high momentum regions in clusters as they may produce large kSZ like effects that would be detectable in high resolution studies of clusters.

It is finally worth noting that in general, each cause of anisotropy in the electron or photon background will have a different angular dependence, as they will have a different angular distribution (i.e., different components under the spherical harmonic decomposition). The expressions derived here allow for the rapid computation of these anisotropic components.
Additionally, while the higher anisotropy behaviour (that is $\ell>2$) may not have as large an impact on observations, these equations nonetheless allow for the size of these effects to be fully computed and analysed. They furthermore allow for the combination of all these anisotropic effects with rSZ and kSZ corrections, to determine how each compare and/or combine. 
In this endeavour, it will be crucial to consider the contributions from various CMB foregrounds \citep[e.g.,][]{Erler2018, Muralidhara2024, Acharya2024} and also the effect of averaging the SZ signal over the cluster \citep{Chluba2013}. With {\tt SZpack}, the modeling of all the SZ-related effects can now be carried out, opening up a range of applications in this respect.

%--------------------------------------------------------------------------------
\subsection{Relation to polarised SZ signals}
%--------------------------------------------------------------------------------
It is also worth discussing the pSZ signals present in clusters as they are an expected feature of anisotropic SZ effects. In particular, it is worth recalling that these relate to polarised signals resulting from quadrupolar components of CMB or electron distributions — compared to the work here discussing effects on the general intensity. Most of the pSZ literature focuses on the CMB anisotropies \citep[e.g.,][]{Sazonov1999} although recent studies have also extended to electron anisotropies \citep[e.g.,][]{Khabibullin2018}. Kinematic effects from cluster rotation \citep{Chluba2001, Chluba2002, Cooray2002, Diego2003} and multiple scatterings can further affect the total polarised cluster morphology.

It is first worth noticing that traditionally these predictions have all been carried out using simplistic models to generate the spectral dependence (i.e., starting from the Kompaneets equation rather than the full cross-section) and therefore provide simplistic models of the eventual spectral shape of the scattering signals. However, using a similar approach to that discussed here, it would be possible to write explicit forms for the $Q$ and $U$ scattering shapes generated by anisotropies in the incoming photons or electrons. This would again allow us to overcome limitations for asymptotic expansions, which have been given in \cite{Itoh2000pSZ}, and furthermore explicitly model the effect of electron anisotropies.

However, we can use the predictions for amplitudes of the pSZ signals to make rough predictions for the scales of anisotropic components to the general intensity. The polarised effects generated by the CMB quadrupole and intrinsic CMB fluctuations are predicted to be around $10^{-8}$ of the CMB temperature \citep[i.e.,][]{Sazonov1999}. These are a similar scale of signal to that expected from the pSZ signals from multiple scatterings within clusters or the higher order kSZ induced pSZ signals. \cite{Khabibullin2018} predicts that magnetically-induced electron anisotropy in shocks could generate a polarised signal of a similar magnitude (around $10$~nK).

Inherently all of the sources of pSZ signals are also sources of anisotropic SZ signals as discussed this far -- caused by quadrupoles in the anisotropies of either the photon or electron field. These predictions for pSZ magnitudes accordingly can give us a prediction for the scale of the quadrupolar contributions to the intensity -- that is broadly at an order of magnitude smaller than the relativistic SZ components. However, this is explicitly only considering the quadrupolar components and not any lower or higher $\ell$ components. Furthermore, the sources of anisotropy are likely to be localised within clusters, and as such higher resolution studies may well lead to more distinct signals even within mere quadrupole considerations.

Finally, polarized SZ signals from non-thermal electron distributions and their anisotropies may open new ways for studying the structure of relativistic outflows and cavities in clusters. Here, not only the density and velocity structure will be important, but also the effect of magnetic fields on the anisotropies in the momentum distributions, warranting studies based on MHD cluster simulations.

%--------------------------------------------------------------------------------
\subsection{Applications beyond the SZ effect}
\label{sec:beyond}
%--------------------------------------------------------------------------------
It is also worth noting that this work explores the analytics relevant for general consideration of anisotropic Compton scattering in the Doppler dominated regime. While here this is considered in the context of the SZ effect and cluster scattering of the CMB, the analytics remain relevant for any process wherein the electron momentum both greatly exceeds the photon momentum ($\pz\gg\omegaz$) and where the photon momentum is not {\it too} high ($\omegaz<0.5$) as has been discussed in \cite{Sarkar2019}. This could be used to further develop new radiative transfer schemes in anisotropic media, or more accurately describe the SZ effect from multiple scatterings \citep[e.g.,][]{Itoh2001, Chluba2014a, Chluba2014b}.

It was also shown that the scattering of the CIB by clusters can in principle be used to study the evolution of this background \citep{Sabyr2022, Acharya2022}. The expressions given here can be used to improve the modeling including non-thermal effects and anisotropies in the distribution functions. This may also have applications in computations of the boosted CIB signals that have been proposed as a new cosmological probe \citep{Maniyar2023} and the radio-SZ effect \citep{Holder2022, Lee2022radio}.

Similarly, the analytic expressions given here can be considered in the context of general anisotropic Compton scattering — similar to how the isotropic SZ kernel \citep{Ensslin2000} can be used in considerations of the general isotropic Compton scattering kernels derived in e.g., \cite{Belmont2009}, \cite{Sarkar2019}. 
Such a generalization would open the path towards applications in the modeling of the evolution of spectral distortion anisotropies in the early Universe \citep{ChlubaTFHI, ChlubaTFHII, ChlubaTFHIII}, which provide a probe of new physics. We leave an exploration of these possibility to future work.

%--------------------------------------------------------------------------------
\section{Conclusion}
\label{sec:conclusion}
%--------------------------------------------------------------------------------
In this work, we obtained the photon and electron scattering kernels including anisotropies in either one of the distributions. Our expressions describe the scattering process in the Doppler-dominated regime and are thus applicable  to the SZ effect and scattering of other backgrounds by energetic electrons. The primary goal was to  demonstrate the validity of the expressions and compare various approaches for several illustrative cases. We also discussed the formation of signals on the underlying electron momentum distribution, highlighting the effects of relativistic non-thermal distributions.

While, the observational outlook for measurements of anisotropic and high-energy non-thermal contributions will require high sensitivity and angular resolution \citep[e.g., CMB-S4, CCAT, Voyage 2050 etc.][]{CCAT2023, Kaustuv2021Voyage, Voyage2050}, they remain an interesting avenue for further study.
Determining the precise details of electrons in the ICM and their impact on observed SZ signals is a critical step in particular for calculating the precise importance of anisotropic electrons to the SZ effect, and high-energy non-thermal electron populations to the SZ signal at both high and low frequencies. 

Anisotropic photon and electron distributions have, heretofore, largely been considered using simplistic arguments, and here we have presented expressions to allow for the precise computation of these anisotropic effects. These have been added as modules to \SZpack, recently rereleased and updated. Furthermore, these equations can provide the basis for a deeper analytic consideration of the pSZ effects, as well as the calculation of these effects under variations to the electron distribution or CMB distortions.

In much the same way, non-thermal high-energy electron distributions have been generally neglected when considering SZ effects. While they likely result in small contributions it may well be of comparable scale to the pSZ signals, albeit with potentially distinctive shapes and characteristics. Further work must be carried out to determine the likely distributions caused by the non-thermal mechanisms in clusters -- i.e., jets and shocks -- and to calculate the prevalence of these non-thermal components in clusters. This will allow for a more detailed understanding of these non-thermal effects in the future.

We have detailed a complementary approach to calculating the kinematic SZ effect to the standard method of distorting the CMB as in \cite{Chluba2012} by instead directly boosting the relativistic Maxwell-Boltzmann distribution of the electrons. This has allowed us to validate each method by determining that they are identical up to the inherent expansion limitations imposed within \cite{Chluba2012}.
We use this as an opportunity to highlight that the new release of \SZpack includes not only new modules for the computation of anisotropic and non-thermal electron and photon distributions, but several previous modules not before publicly released. It also features a fully-integrated {\tt Python} wrapper to allow for the easy utility of \SZpack in calculating arbitrary SZ signals for theoretical modelling, forecasting and direct analysis of data.

With the many ongoing and upcoming experiments \citep[e.g.,][]{ACT2020Choi, TheSimonsObservatoryCollaboration2018, CCAT2023, Kaustuv2021Voyage} promising higher accuracy and higher resolution imaging of galaxy clusters, anisotropic effects in clusters will become increasingly relevant. As such, further work exploring the magnitude and detectability of these signals in physically-motivated models of clusters is an avenue for future exploration.

%----------------------------------------------------------------------------

{\small
\section*{Acknowledgements}
This work was supported by the ERC Consolidator Grant {\it CMBSPEC} (No.~725456).
EL was also supported by the Royal Society on grant No.~RGF/EA/180053.
JC was furthermore supported by the Royal Society as a Royal Society University Research Fellow at the University of Manchester, UK (No.~URF/R/191023).
}
%--------------------------------------------------------------------------------
\section*{Data Availability}
%--------------------------------------------------------------------------------
\SZpack has been used for the generation of all the figures in this work. An updated and revised form of \SZpack can be found at \href{https://github.com/CMBSPEC/SZpack}{github.com/CMBSPEC/SZpack}.

%--------------------------------------------------------------------------------
{\small
\bibliographystyle{JHEP}
\bibliography{HighEnergySZ}
}

\newpage
\appendix

%----------------------------------------------------------------------------
\section{Details of the derivations}
\label{app:derivation}
%----------------------------------------------------------------------------
Some consideration is required in the derivation of the anisotropic photon and electron kernels, which will be presented here. Assuming that one of the distributions is isotropic, the required kernels can be defined as
%---------------------------------------------------------
\begin{equation} \begin{split}
\label{eq:K_ints}
    K^{m0}_{\ell0}(t,\pz) &\equiv \int  \frac{\D{\mu_0}\D{\musc}\D{\phi_{\rm sc}}}{\sqrt{4\pi}\,\sigT} \frac{\D{\phi_0}}{\D{t}}\frac{\D{\sigma}}{\D{\Omega}} Y^m_\ell(\musc, \phi_{\rm sc})\\
    K^{0m}_{0\ell}(t,\pz) &\equiv \int \frac{\D{\mu_0}\D{\musc}\D{\phi_0}}{\sqrt{4\pi}\,\sigT} \frac{\D{\phi_{\rm sc}}}{\D{t}}  \frac{\D{\sigma}}{\D{\Omega}} Y^m_\ell(\muz, \phi_0).
\end{split} \end{equation}
%---------------------------------------------------------
where $\D{\sigma}/\D{\Omega}$ is a function of $(t, \pz, \mu_0, \mu', \mu_{\rm sc})$ given by
%----------------
\begin{equation}
\label{eq:App_kerdef2_deff}
\frac{\text{d}\sigma}{\text{d}\Omega} \approx  \frac{3\sigT}{16\pi}\,\frac{t^2\bar{X}(\pz, \mu_0, \mu', \mu_{\rm sc})}{\gamma^2_0(1-\beta_0\mu_0)}.
\end{equation}
%----------------
It is worth mentioning that $\bar{X}$ is symmetric in $\mu_0 \leftrightarrow \mu'$, as, in the SZ case, it follows the form:
\begin{equation}
\nonumber
\bar{X} = 2\left[1 - \frac{(1-\mu_{\rm{sc}})}{\gamma_0^2(1-\beta_0\mu_0)(1-\beta_0\mu')} + \frac{(1-\mu_{\rm{sc}})^2}{2\gamma_0^4(1-\beta_0\mu_0)^2(1-\beta_0\mu')^2}\right]. 
\end{equation}
The dependence on $\phi_{\rm sc}$ and $\phi_0$ implicitly enters the problem through $$\mu'=\mu_0\mu_{\rm sc}+\cos(\phi_0-\phi_{\rm sc})\sqrt{(1-\muz^2)(1-\musc^2)}.$$ By recalling that $t=(1-\beta_0\mu_0)/(1-\beta_0\mu')$, we can then solve for $\xi=\cos(\phi_0-\phi_{\rm sc})$. This implies
%--------------------------------------------------------------------------------
\begin{equation} \begin{split}
    \xi &= \frac{1-t^{-1}+\betaz\muz(t^{-1}-\musc)}{\betaz\sqrt{(1-\muz^2)(1-\musc^2)}}
    \\
    \frac{\D{\phi_0}}{\D{t}}&=-\frac{\D{\phi_{\rm sc}}}{\D{t}}=-\frac{\D{\xi}/\D{t}}{
    \sqrt{1-\xi^2}} \nonumber 
    =-\frac{1-\betaz\muz}{t^2\sqrt{\betaz^2(1-\muz^2)(1-\musc^2)-[1-t^{-1}+\betaz\muz(t^{-1}-\musc)]^2}}.
\end{split} \end{equation}
%--------------------------------------------------------------------------------
Put together, we then have \citep[see also][Eq.~A10 once dropping small terms]{Sarkar2019}
%---------------------------------------------------------
\begin{align}
    \mathcal{J}&=\Bigg|\frac{\D{\phi_0}}{\D{t}}\Bigg|\,\frac{\D{\sigma}}{\sigT\D{\Omega}}
    =
    \frac{3}{8\pi \gamma_0 p_0}
    \frac{1 - \frac{t(1-\mu_{\rm{sc}})}{\gamma_0^2 (1-\beta_0\mu_0)^2} + \frac{t^2(1-\mu_{\rm{sc}})^2}{2\gamma_0^4(1-\beta_0\muz)^4}}{\sqrt{(1-\muz^2)(1-\musc^2)-\frac{[t-1+\betaz\muz(1-t\musc)]^2}{\beta_0^2 t^2}}}.
\end{align} 
%---------------------------------------------------------
This implies that in the integrals Eq.~\eqref{eq:K_ints} the $\phi_0$ and $\phi_{\rm sc}$ dependence now only enters through the spherical harmonic factors.  This implies that only the kernels for $m=0$ matter, reducing the problem to
%---------------------------------------------------------
\begin{equation} \begin{split}
\label{eq:K_ints_simp}
    K^\gamma_{\ell}=\frac{K^{00}_{\ell0}}{\sqrt{2\ell +1}}&\equiv \frac{1}{2}\int \D{\mu_0}\D{\musc} \mathcal{J}(\beta_0, t, \mu_0, \musc)\,P_\ell(\musc)
    \\
    K^{\rm e}_{\ell}=\frac{K^{00}_{0 \ell}}{\sqrt{2\ell +1}} &\equiv \frac{1}{2}\int \D{\mu_0}\D{\musc} \mathcal{J}(\beta_0, t, \mu_0, \musc)\,P_\ell(\muz).
    \end{split} 
    \end{equation}
%---------------------------------------------------------
We note that the range of the integrals still has to be specified.
Also, when considering {\it simultaneous} electron and photon anisotropy, the transformation to $\phi_0$, firstly, leads to the $\phi_{\rm sc}$ 
 dependence $\expf{i(m+m')\phi_{\rm sc}}$, so the $m=-m'$ terms still contribute, and further leads to significant complications to the rest of the integration process.

\subsection{Photon scattering kernels}
%-----------
The range of $\mu_0$ and $\mu_{\rm sc}$ follows from the condition that the argument of the root has to remain real. For the photon kernel it is best to first integrate over $\mu_0$ and then $\mu_{\rm sc}$, which requires
%--------------------------------------------------------------------------------
\begin{equation} \begin{split}
    \mu_0^{\pm} &= \frac{(1-t)(1-t\musc)
    %\\&\qqquad
    \pm
    \betaz t 
    \sqrt{2t\left(1-\musc^2\right) (\mu_{\rm sc}^{\rm cr} -\musc )}}{2\betaz t \left(\frac{1+t^2}{2t} -\musc\right)},
    \\
    \mu_{\rm sc}^+&=\mu_{\rm sc}^{\rm cr},\quad
    \mu_{\rm sc}^-=-1, \quad
    \mu_{\rm sc}^{\rm cr}=1-\frac{(1-t)^2}{2\pz^2 t}
    %=1-\frac{\alpha_-}{p_0^2}
\end{split} \end{equation}
%--------------------------------------------------------------------------------
We note that $(1+t^2)/(2t)>1$ for all $t$ and $\mu_{\rm sc}^{\rm cr}=1$ for $t=1$. In addition, $-1\leq \mu_{\rm sc}^{\rm cr}\leq 1$ for $$t^-=\frac{1-\betaz}{1+\betaz}\leq t \leq \frac{1+\betaz}{1-\betaz}=t^+,$$ which naturally is the maximally allowed scattering range. With this we can rewrite $\mathcal{J}$ as
%---------------------------------------------------------
\begin{align}
    \mathcal{J}_\gamma&=
    \frac{3 t}{8\pi \gammaz p_0}
    \frac{1 - \frac{t(1-\mu_{\rm{sc}})}{\gamma_0^2 (1-\beta_0\mu_0)^2} + \frac{t^2(1-\mu_{\rm{sc}})^2}{2\gamma_0^4(1-\beta_0\muz)^4}}{\sqrt{\left(1+t^2-2t\musc\right)(\mu_0^{+}-\mu_0)(\mu_0-\mu_0^{-})}},
\end{align} 
%---------------------------------------------------------
where the subscript '$\gamma$' clarifies that this applies to the photon kernels.
Carrying out the integral over $\mu_0$ then yields
%---------------------------------------------------------
\begin{align}
    &\mathcal{I}_\gamma=
    \int_{\mu_0^-}^{\mu_0^+}\frac{\D{\mu_0}}{2}\mathcal{J}_\gamma
    \nonumber \\[-3mm]
    &\quad =
    \frac{3}{8 \gammaz p_0}
    \frac{t}{\sqrt{1+t^2-2t\musc}}\Bigg\{
    1 - \frac{t(1-\mu_{\rm{sc}})\left[1-\beta_0\frac{\mu_0^-+\mu_0^+}{2}\right]}{\gamma_0^2 ([1-\beta_0\mu_0^-][1-\beta_0\mu_0^+])^{3/2}}
    \nonumber \\ \nonumber
    &\qqquad \times \left(1-
    \frac{t(1-\mu_{\rm{sc}})
    \left[1-\beta_0(\mu_0^-+\mu_0^+)+\frac{5}{8}\beta_0^2(\mu_0^-+\mu_0^+)^2-\frac{3}{2}\beta_0^2\mu_0^-\mu_0^+\right]}{2\gamma_0^2 ([1-\beta_0\mu_0^-][1-\beta_0\mu_0^+])^2}\right)\Bigg\}
    \\
    &\quad =\frac{3}{8 \gammaz p_0}
    \Bigg\{
    \frac{t
    }{\sqrt{1+t^2-2t \musc}} - \frac{\gamma_0(1+t)\sqrt{1-\mu_{\rm{sc}}}}{[2+p_0^2 (1-\mu_{\rm{sc}})]^{3/2}}\,\times 
    \nonumber \\ \nonumber
    &\qquad\qquad \qquad 
   \left(1+
    \frac{3
    \left(\alpha_- + 1 -\mu_{\rm{sc}}\right)}{2[2+p_0^2 (1-\mu_{\rm{sc}})]}
    -
    \frac{5\gamma_0^2\alpha_+(1-\mu_{\rm{sc}})}{2[2+p_0^2 (1-\mu_{\rm{sc}})]^2}\right)\Bigg\},
\end{align} 
%---------------------------------------------------------
with $\alpha_\pm = (1\pm t)^2/[2t]$ and where we used the identities
%-------------------------------
\begin{align} 
\mu_0^-+\mu_0^+&=\frac{2(1-t)(1-t\musc)}{\betaz \left(1+t^2 - 2t\musc\right)}
\\
(1-\beta_0\mu_0^-)(1-\beta_0\mu_0^+)&=\frac{t^2(1-\mu_{\rm{sc}})[2-\beta_0^2 (1+\mu_{\rm{sc}})]}{\left(1+t^2-2t\musc\right)}.
\end{align} 
%-------------------------------
The first term in $\mathcal{I}_\gamma$ can principally be directly rewritten in terms of the Legendre polynomials. However, since the integration range is limited, this does not actually simplify the computation. However, by using $\kappa_{\rm sc}=1-\mu_{\rm sc}$ for this term and $y=\sqrt{1+\frac{1}{2}\pz^2 \kappa_{\rm sc}}$ we can recast $\mathcal{I}_\gamma$ into the more compact form
%---------------------------------------------------------
\begin{align}
\label{eq:def_I_Int}
    \mathcal{I}_\gamma
    &=\frac{3}{8 p_0}
    \Bigg\{\frac{f_0}{\gammaz } + \frac{(1+t)}{4t}
    \Bigg(5\gamma_0^2 (1+t)^2 \,g_0
    \nonumber -\left[12\gamma_0^2 t+(3+2 p_0^2)(1+t)^2\right] g_1+ 4t\,(3+2 p_0^2)\,g_2
    \Bigg)\,
    \Bigg\}
    \nonumber 
    \\[0mm]
f_0&=   
\frac{t
    }{\sqrt{1+t^2-2t \musc}} = \frac{\sqrt{t/2}
    }{\sqrt{\alpha_- +  \kappa_{\rm sc} }}, \qquad g_m=-\frac{\sqrt{y^2-1}\,y^{2m}}{4 p_0^3 y^7}
\end{align} 
%---------------------------------------------------------
By using $P_\ell(\musc) = \sum_{k=0}^\ell\binom{\ell}{k}\binom{\ell+k}{k} \left(\frac{\musc-1}{2}\right)^k$, we encounter the integrals of the form $$G_k=\int^2_{\frac{\alpha_-}{p_0^2}} \left(\frac{-\kappa_{\rm sc}}{2}\right)^k \frac{f_0}{\gamma_0}\,\D{\kappa_{\rm sc}},$$ which can be readily carried out yielding
%---------------------------------------------------------
\begin{align}
G_k&=\frac{|1-t|^{2k+1}}{\gamma_0 (4 t)^k}
\Bigg[
\frac{1+t}{|1-t|}\, {}_2F_1\left(
-k,\frac{1}{2},\frac{3}{2},\frac{(1+t)^2}{|1-t|^2}
\right)
-\frac{\gamma_0 }{p_0} {}_2F_1\left(-k,\frac{1}{2},\frac{3}{2},\frac{\gamma_0^2}{p_0^2}\right)
\Bigg]
\nonumber \\ \nonumber
&=\sum_{m=0}^k \frac{(-1)^m}{2m+1} \binom{k}{m}
\,\frac{|1-t|^{2k-2m}}{4^k t^k} 
\left((1+t)^{2m+1}-\frac{\gamma_0^{2m+1}}{p_0^{2m+1}}\,|1-t|^{2m+1}\right).
\end{align} 
%---------------------------------------------------------
This explains the corresponding term in Eq.~\eqref{eq:K_gamma_gen_res}. 

For the other integrals we integrate over $y$ instead of $\kappa_{\rm sc}$. Carrying out the transformations and defining the function $h_m^{(k)}$ as given by Eq.~\eqref{eq:def_h} one then has
%---------------------------------------------------------
\begin{align}
H^{(k)}_m&=
\int^2_{\frac{\alpha_-}{p_0^2}} \left(\frac{-\kappa_{\rm sc}}{2}\right)^k g_m\,\D{\kappa_{\rm sc}}
\equiv 
\int_{\gamma_0}^{\sqrt{1+\frac{\alpha_-}{2}}}
\frac{\sqrt{y^2-1}\,(1-y^2)^k\,y^m}{p_0^{2k+5} y^6} \D{y}
=\frac{h_{2m}^{(k)}}{p_0^{2k+5}}.\end{align} 
%---------------------------------------------------------
The integrals can again be expressed using hypergeometric functions. Collecting terms then yields the expressions in Eq.~\eqref{eq:K_gamma_gen_res}.

\subsection{Recurrence relations for photon kernel}
%-----------
We can also write recurrence relations for the full integrals. Defining the auxillary functions %---------------------------------------------------------
\begin{align}
\mathcal{H}_{\ell}^{(m)}&=\int_{\gamma_0}^{\sqrt{1+\frac{(1-t)^2}{4t}}} \frac{\sqrt{y^2-1}\,y^{2m}}{p_0^5\,y^6}P_\ell\left(1-\frac{2}{p_0^2}[y^2-1]\right)\,\D{y}
\end{align} 
%---------------------------------------------------------
and using the recurrence relations for $P_\ell$ we find
%---------------------------------------------------------
\begin{align}
\mathcal{H}_{\ell+1}^{(m)}
&=\frac{(2\ell +1)}{\ell+1}
\left[
\frac{2+p_0^2}{p_0^2}
\,\mathcal{H}_{\ell}^{(m)}
-\frac{2}{p_0^2}\,\mathcal{H}_{\ell}^{(m+1)}
\right]
-\frac{\ell}{\ell+1}\,\mathcal{H}_{\ell-1}^{(m)}.
\end{align} 
%---------------------------------------------------------
This recurrence can be started knowing all $\ell=0$ results and using that $\mathcal{H}_{-1}^{(m)}=0$. 
This yields
%---------------------------------------------------------
\begin{subequations}
\begin{align}
    K^\gamma_\ell
    &=\frac{3}{8 p_0}
    \bigg(\mathcal{G}_\ell+\mathcal{H}_\ell \bigg)
\\
\mathcal{G}_\ell&=
\int^2_{\frac{\alpha_-}{p_0^2}} 
f_0\, P_\ell \big(1-\kappa_{\rm sc}\big)\,\D{\kappa_{\rm sc}}
=
\sum_{k=0}^\ell\binom{\ell}{k}\binom{\ell+k}{k} \,G_k
\\
\mathcal{H}_\ell&=\frac{(1+t)}{4t}
    \Bigg\{5\gamma_0^2 (1+t)^2 \,\mathcal{H}^{(0)}_\ell
    -\left[12\gamma_0^2 t+(3+2 p_0^2)(1+t)^2\right] \mathcal{H}^{(1)}_\ell
     + 4t\,(3+2 p_0^2)\,\mathcal{H}^{(2)}_\ell
    \Bigg\}\,.
\end{align} 
\end{subequations}
%---------------------------------------------------------
By using the recurrence relations, one can also eliminate the explicit dependencies on $\mathcal{H}^{(m)}_\ell$ for $m>0$, however, this does not provide any additional benefit. We also note that $$\mathcal{H}^{(m)}_\ell = \sum_{k=0}^\ell\binom{\ell}{k}\binom{\ell+k}{k} \,H_k^{(m)}=\sum_{k=0}^\ell\binom{\ell}{k}\binom{\ell+k}{k} \,\frac{h_{2m}^{(k)}}{p_0^{2k+5}}$$ per definition, which completes the analysis for the photon kernels.

\subsection{Electron scattering kernels}
\label{app:derivation_e}
%-----------
To derive the result for anisotropic electrons, we first integrate over $\mu_{\rm sc}$ and then $\mu_0$. In practice we will use $\lambda_0=1-\betaz \muz$ instead of $\mu_0$. This implies the integration bounds
%--------------------------------------------------------------------------------
\begin{subequations}
\begin{align}
    \mu_{\rm sc}^{\pm} &= \frac{
    \gamma_0 \mu_0[t-\lambda_0]
    %\\&\qqquad
    \pm
    \sqrt{\left(1-\muz^2\right) \left[t-\frac{\lambda_0}{1+\beta_0}\right]\left[\frac{\lambda_0}{1-\beta_0}-t\right]}}{\pz t},
\\
\lambda_0^{-}&=
\begin{cases}
& 1-\betaz \qquad {\rm for} \;t\leq 1
\\
& t(1-\betaz)  
\quad\, {\rm else}
\end{cases}
, \quad
\lambda_0^{+}=
\begin{cases}
& t(1+\betaz) \quad {\rm for} \;t\leq 1
\\
& 1+\betaz  
\qquad{\rm else}
\end{cases}
\end{align}
\end{subequations}
%--------------------------------------------------------------------------------
For the integral over $\musc$ this then yield
%---------------------------------------------------------
\begin{align}
    &\mathcal{I}_{\rm e}=
    \int_{\musc^-}^{\musc^+}\frac{\D{\musc}}{2}\mathcal{J}
    \nonumber \\
    &\quad =\frac{3}{32 \gammaz \pz^5}
    \Bigg\{3+4\pz^2+4\pz^4 
    +\frac{3t^2}{\gammaz^4\lambda_0^4}
    -\frac{6t(1+t)}{\gammaz^2\lambda_0^3}
    \nonumber \\ \nonumber
    &\qquad\qquad\qquad 
   +\frac{(3+2\pz^2)(1+t^2) + 12 \gammaz^2 t}{\gammaz^2\lambda_0^2}
   -\frac{2(3+2\pz^2)(1+t)}{\lambda_0}
   \Bigg\}.
\end{align} 
%---------------------------------------------------------
Again inserting $P_\ell(\musc) = \sum_{k=0}^\ell\binom{\ell}{k}\binom{\ell+k}{k} \left(\frac{\musc-1}{2}\right)^k$, we encounter the following integrals
%---------------------------------------------------------
\begin{align}
X^{(k)}_m&=
\int^{\lambda_0^+}_{\lambda_0^-}  
\left(\frac{\musc-1}{2}\right)^k
\frac{\D{\lambda_0}}{\betaz \gamma_0^m \lambda_0^m}
\nonumber
%\\&
=
\frac{(1-\betaz)^{k+1-m} }{2^k \betaz^{k+1}\,\gamma_0^m}
\int^{\tau_0^+}_{\tau_0^-} 
\left(1-\tau_0\right)^k \frac{\D{\tau_0}}{\tau_0^m}
\end{align} 
%---------------------------------------------------------
with $\tau_0=\lambda_0/(1-\betaz)$. The remaining integral can be solved using the binomial formula. As an intermediate step it is useful to solve the integrals
$g^{(\alpha)}=\int^{\tau_0^+}_{\tau_0^-} 
\tau_0^{\alpha-1} \D{\tau_0}$, which yields
%---------------------------------------------------------
\begin{align}
g^{(\alpha)}
&=
\begin{cases}
-|\log(t)|+2\sinh^{-1}(\pz) &\alpha = 0
\\[2mm]
\displaystyle 
\frac{(t^\alpha+1)(t_+^\alpha-1)
    -{\rm sign}(t-1)(t^\alpha-1)(t_+^\alpha+1)}{2 \alpha}
    &{\rm otherwise}
\end{cases}
\end{align} 
%---------------------------------------------------------
We note that $2\sinh^{-1}(\pz)\equiv \log(t+)$ and that for $\alpha\neq 0$ one has $g^{(\alpha)}=[(t\, t_+)^\alpha-1]/\alpha$ for $t_-\leq t\leq 1$ and $g^{(\alpha)}=[t_+^\alpha-t^\alpha]/\alpha$ for $1\leq t\leq t_+$. With this we then find
%--------------------------------------------------------------------------------
\begin{align} 
\label{eq:e_kernel}
    K^{\rm e}_{\ell}(t,\pz) &=\sum_{k=0}^{\ell}\binom{\ell}{k}\binom{\ell+k}{k}\Bigg[\frac{3}{32 \gammaz \pz^5} \Big(3 t^2 X_4^{(k)}-6\gammaz \, t(t+1) \, X_3^{(k)}+\left[(3+2\pz^2)(1+t^2) + 12 \gammaz^2 t\right] X_2^{(k)}
    \nonumber\\\nonumber
    &\qquad\qquad-2 \gammaz (3+2\pz^2)\,(1+t)\,X_1^{(k)}+(3+4\pz+4\pz^2)\,X_0^{(k)}\Big)\Bigg]; 
    \\
    X_m^{(k)} &= \frac{(\gammaz-\pz)^{k+1-m}}{2^k \pz^{k+1}} \sum_{n=0}^{k}\,(-1)^{n}\,\binom{k}{n} \;g^{(n+1-m)}.
\end{align}
%--------------------------------------------------------------------------------
For $\ell =0$ This agrees with the photon scattering kernel, as expected.

%--------------------------------------------------------------------------------
\section{Numerical stability of the kernels}
\label{app:numericalStability}

Here we quantitatively describe the numerical stability of the analytic forms of the anisotropic photon and electron kernels. While in principle it may be possible for an arbitrary $\ell$ to determine a kernel which is numerically stable to a certain degree of accuracy, for kernels derived for a general $\ell$, as we have presented in Section~\ref{sec:Theory}, this is substantially more complex.

As a result, within \SZpack and for the figures displayed in this work, we use the analytic expressions where they are numerically stable, and revert to a full numerical integration where the analytic forms become unstable. For these purposes we define the error in an analytic form as 
%--------------------------------------------------------------------------------
\begin{equation} \label{eqn:stability}
    \epsilon(\pz) = \frac{\sum_{t=t_i} |K_{\rm analytic}(t,\pz)-K_{\rm integrated}(t,\pz)|}{N_{\rm{max}}(|K_{\rm integrated}(t_i,\pz)|)}.
\end{equation}
%--------------------------------------------------------------------------------
Here $N=101$, the number of logarithmically spaced data points we summed over, between the two t limits of $t_\pm= (\gammaz\pm\pz)/(\gammaz\mp\pz)$. $K_{\rm analytic}$ and $K_{\rm integrated}$ refer respectively to the analytic form a given kernel or the numerically integrated kernel (where that kernel may be photon or electron, and at any given $\ell)$. The notation $\rm{max}(|x_i|)$ indicates that the maximal absolute value of the $N$ calculated was used here.
Due to the intrinsic variability as $\pz$ varied, we then took the median value of [$\epsilon(0.998\bar{p})$, $\epsilon(0.999\bar{p})$, $\epsilon(1.000\bar{p})$, $\epsilon(1.001\bar{p})$, $\epsilon(1.002\bar{p})$]. 

\begin{figure}[H]
    \centering
    \includegraphics[width=0.58\columnwidth]{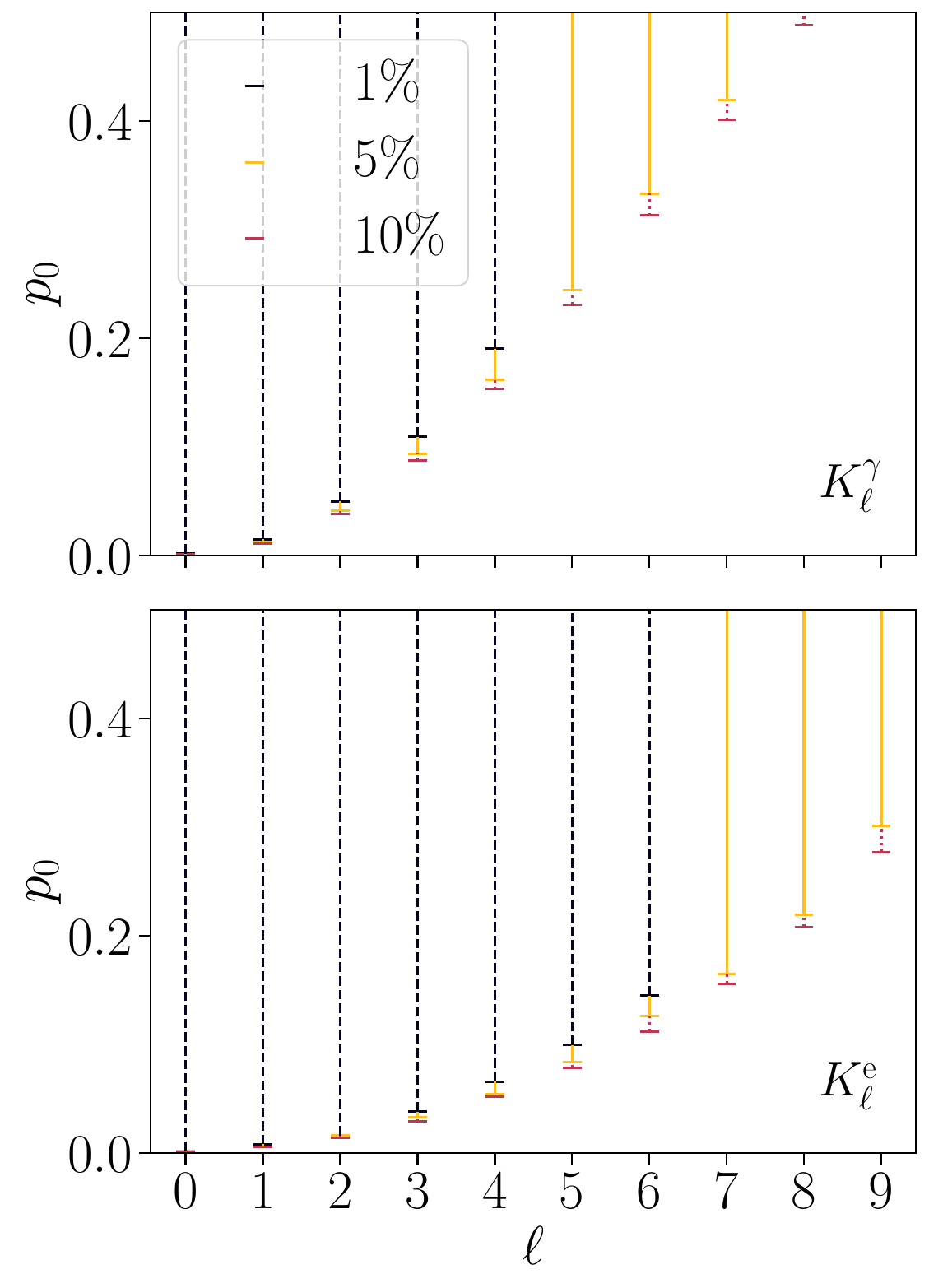}
    \caption[Numerical Stability]{The stability regions of each multipole ($\ell<10$) for the photon and electron kernel (top and bottom panels respectively). Displayed here are the regions within which the error (as determined by Eq.~\ref{eqn:stability}) of the analytic form is less than the indicated value (0.3\%, 1\% and 10\%). }
    \label{fig:A_numerical_stability}
\end{figure}

We then explored the numerical stability of each kernel through three criteria error values, of $\epsilon(\bar{p})= 0.1$, 0.05 and 0.01. We then found the lowest $\pz$ at which these conditions appeared to be stably met. This gave us conditions on $\pz$ that we will refer to as 10\%, 5\% and 1\%. These conditions for all $\ell\leq9$ for both the photon and electron kernels are displayed in Fig.~\ref{fig:A_numerical_stability}.

We can immediately see that for both kernels the instabilities become more prevalent for higher multipoles (that is, higher $\ell$) -- which can be predicted from the analytic forms displayed in Section~\ref{sec:Theory}. Each kernel contain inverse factors of $\pz$ to higher orders with higher multipoles -- and, as such, as $\pz\rightarrow0$, errors from these calculations become increasingly prevalent.\footnote{It is likely that much of this instability could be avoided on a multipole by multipole term by expanding ${\rm arcsinh}(\pz)$ about $\pz=0$ and gathering terms appropriately.}

At the higher multipoles, the inherent stability of the metric becomes harder to ascertain. That is, since the magnitude of the kernel drops with increasing $\ell$, the denominator in the metric becomes increasingly small, and begins to dominate the observed kernel. As such, the 1\% metric ceased to be achievable, and so is not present here. 

It is also evident that the analytic form of the electron kernel presented here is more stable than the analytic form of the photon kernel. This can be also explained, to an extent, from considering the forms themselves -- the electron kernel form is simplified and the terms are more directly condensed. The photon multipole kernel, in contrast, as written contains a complicated mixing of terms leading to far trickier and less accurate cancelling of terms.

Nonetheless, to manage these issues, in \SZpack a stable form of the kernels has been introduced which at low $p$, that is, less than the 1\% criteria, calculates the the kernel through a full numerical integration, and above this criterion uses the analytic forms as presented above. For $\ell\geq10$, this stable form uses always the numerical integration, sacrificing speed for numerical accuracy. 

%--------------------------------------------------------------------------------
%--------------------------------------------------------------------------------

%--------------------------------------------------------------------------------
\section{Comparison of the kSZ approaches}
\label{app:kSZ_comp}

Since the CNSN method used in e.g., Fig.~\ref{fig:5b_boost_vs CNSN} is curtailed to various orders of $\betac$ and is difficult to compare directly on a multipole to multipole basis with the CNSN numerical methods.

However, when considering the signal as a whole, it is possible to instead compare this electron-boosted method in the CMB frame, with a photon-boosted method in the cluster frame. Where one must only ensure that the results from the cluster frame are correctly converted into the CMB frame -- as described in detail in CNSN (see e.g., their section 4.2).
Hence, we must note that the photon distributions are boosted as, 
%-------------------------------------------------
\begin{equation*} 
    n(x') = \frac{\expf{-\gammac x' (1+\betac\muc')}}{1- \expf{-\gammac x' (1+\betac\muc')}}, \quad
    n(x) = \frac{\expf{-\gammac x (1+\betac\muc)}}{1- \expf{-\gammac x (1+\betac\muc)}}, \quad
    \muc' = \muc\mup +\cos{(-\phi')}\sqrt{(1-\muc^2)(1-\mup^2)}.
\end{equation*}
%-------------------------------------------------
It is worth recalling here that $\muc$ is the cosine of the angle between the cluster velocity and the line-of-sight and $\mup$ is the cosine of the angle between the incoming and outgoing photon. $\phi'$ is the azimuthal angle between the incoming and outgoing photons.
This can then be integrated with a full 5D integration within \SZpack and gives us a reference to which we can compare the boosted-electron method described in Section~\ref{sec:kSZ_electrons}.

%--------------------------------------------------------------------------------
\begin{figure}[H]
    \centering
    \includegraphics[width=0.65\columnwidth]{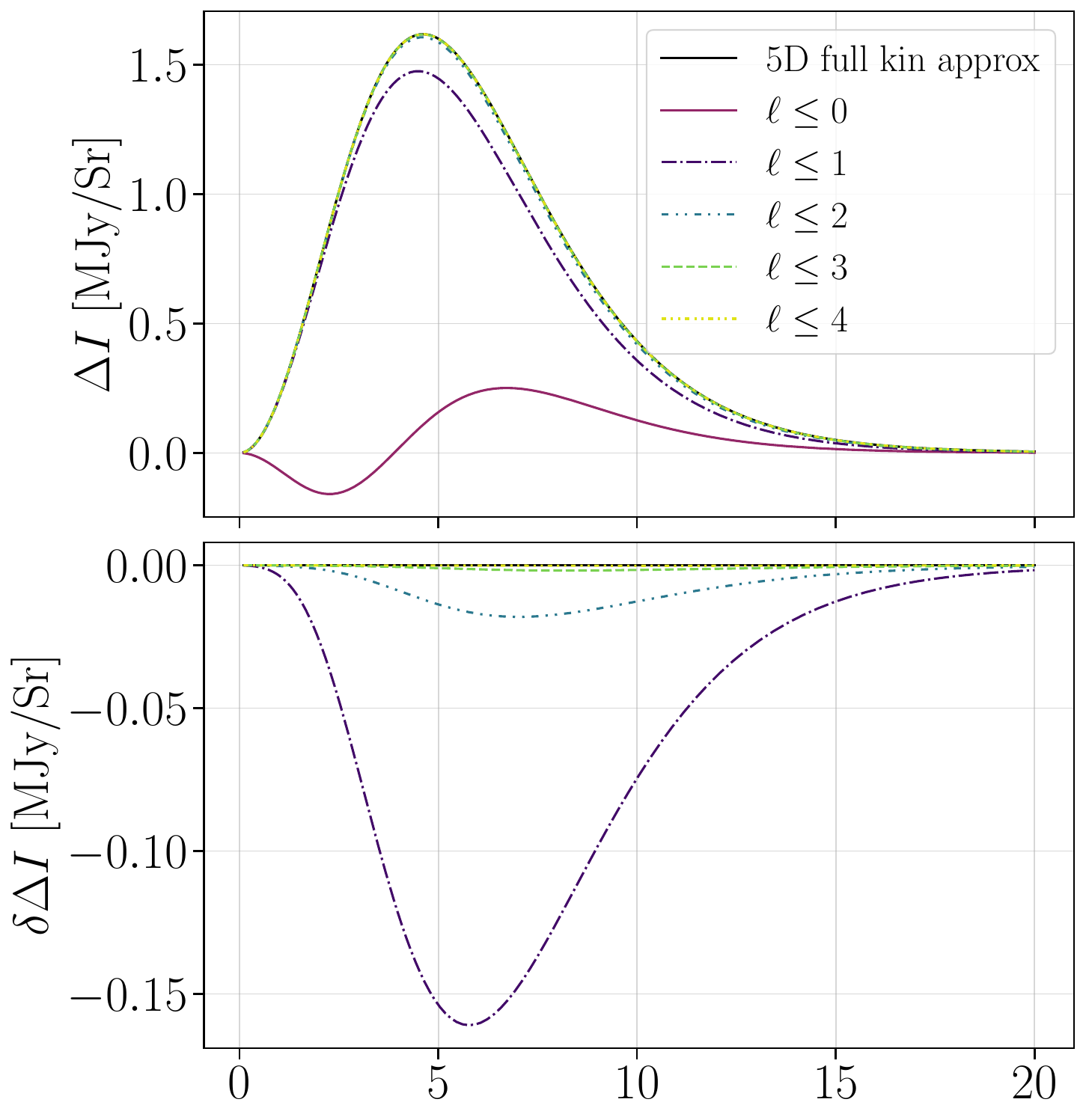}
    \caption[Comparison to of anisotropic and 5D kSZ approaches]{A comparison of the anisotropic electron-boosted approach to the kinematic SZ effect with a 5D integral over the photon-boosted approach described in CNSN -- here labelled `5D full kin approx'. The anisotropic are the sum described in Eq.~\eqref{eqn:boostMB} curtailed to increasing $\ell$. The top panel shows the distortion and the bottom panel shows the difference between each signal and the 5D full kin approach. Here $\muc=1.0$, $\betac = 0.1$, $\Te=5$~keV and $y=10^{-4}$. }
    \label{fig:D_Anisotropic_5D_diff}
\end{figure}
%--------------------------------------------------------------------------------

The results of this comparison are displayed in Fig.~\ref{fig:D_Anisotropic_5D_diff}, where we have shown the impact of including higher multipoles in the anisotropic electron expansion on the convergence of the kSZ signal for $\betac=0.1$. It is immediately clear that the dipole term ($\ell=1$) contains the largest correction to the tSZ signal, and allows for a signal that agrees with the 5D calculation to $\simeq10$\% with the strongest diversion around the peak of the kSZ signal.

With the inclusion of the quadrupole term ($\ell=2$), this agreement increases to $\simeq1$\%, and increases again with the inclusion of the octupolar and hexadecapolar terms ($\ell=3$ and $4$) each with an decrease in difference of around a factor of 10. Indeed, by this point, the agreement is likely limited as much by the numerical convergence of the integration routines as it is by the marginal gains added at this cluster velocity and temperature in including higher multipoles.

It is thus evident that this method reaches convergence with the `true' kinematic signal (as determined through the detailed 5D integral in the cluster frame) to a high degree of accuracy with a comparatively small number of multipoles. This indeed matches well with the conclusions found within CNSN on the convergence of spherical harmonic expansions of the boosted photons within the cluster frame.
This also allows us to verify that the anisotropic electron method described in this work is correct in its formulation.

%--------------------------------------------------------------------------------
\section[Velocity expansion of the kinematic SZ formulations]{$\betac$ expansion of kSZ formulations}
\label{app:kSZ_betacexp}

An obvious distinction between the two formulations of the kinematic SZ effects described in Sections~\ref{sec:kSZ_photons} and \ref{sec:kSZ_electrons} is the curtailing of the expansion to second order in $\betac$ in the CNSN formulation compared to the electron anisotropy method detailed here. As such, we here confirm that this distinction is not the cause of the deviation between the two forms of the kinematic SZ signal. 

\begin{figure}[H]
    \centering
    \includegraphics[width=0.55\columnwidth]{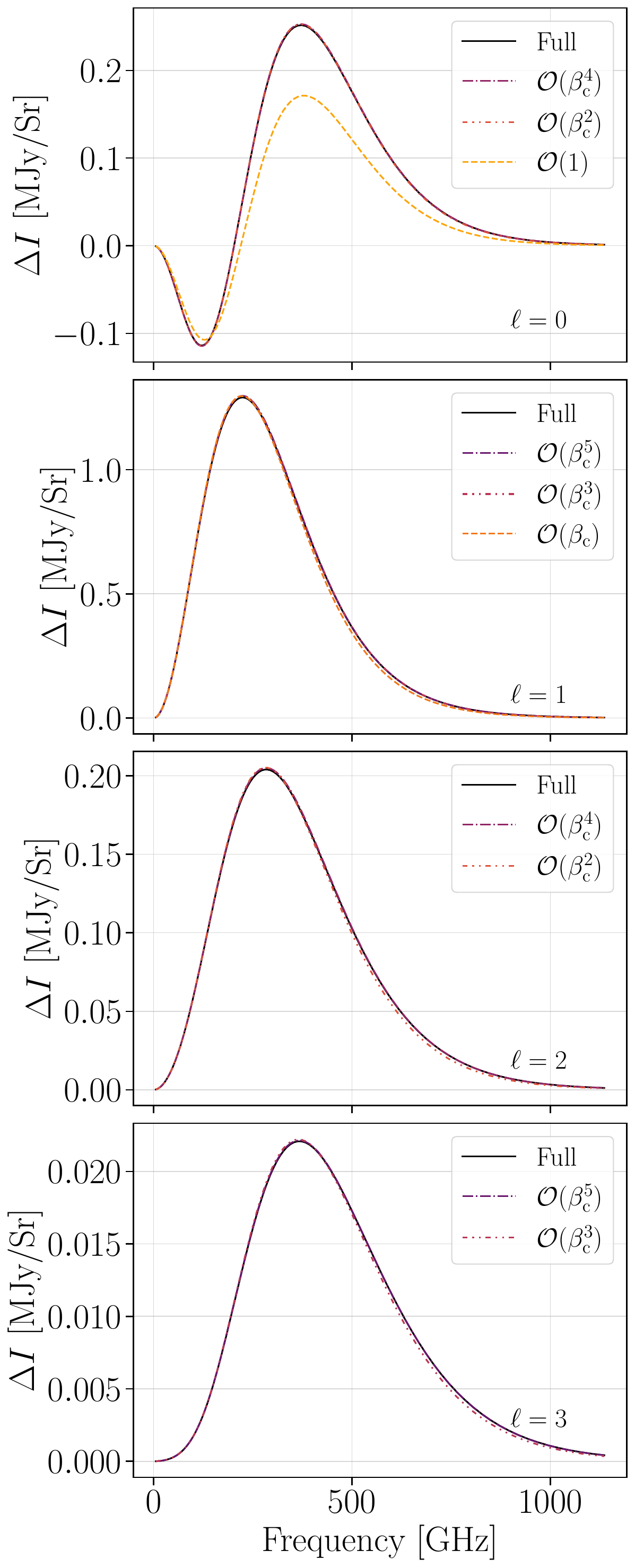}
    \caption[Velocity expansion of boosted SZ]{The convergence of the SZ signals from a boosted Maxwell-J\"uttner distribution for the first four angular multipoles with increasing orders of $\betac$. Here, $\betac=0.1$, $\muc=1.0$, $\thetae=5$~keV and $y=10^{-4}$.}
    \label{fig:C_betac_exp_electron_boost}
\end{figure}

To do so, it is necessary to expand the anisotropic forms of the boosted Maxwell-Boltzmann distributions, $f_{\rm boost}^{(\ell)}$ (Eq.~\ref{eqn:boostMB}) in terms of $\betac$. This yields, for $\ell \leq 3$,
%--------------------------------------------------------------------------------
\begin{equation} \begin{split}
    f_{\rm boost}^{(0)}(p) &= f_{\rm th}(p) \Bigg[1+\frac{\betac^2}{6\thetae^2}(p^2-3\gamma\thetae)+\frac{\betac^4}{120\thetae^4}\left(p^4-10\thetae\gamma p^2+5\thetae^2(7p^2+3-9\gamma\thetae)\right) +\mathcal{O}(\betac^6) \Bigg] 
    \\
    f_{\rm boost}^{(1)}(p) &= f_{\rm th}(p) P_1(\mu)\,\frac{\betac p}{\thetae}\,\Bigg[1+\frac{\betac^2}{10\thetae^2}\left(p^2+5\thetae(\thetae-\gamma)\right)\\
    &\qquad +\frac{\betac^4}{280\thetae^4}\Bigg(p^4-14\thetae\gamma p^2+7\thetae^2\Big[11p^2+5(1-5\thetae\gamma+3\thetae^2)\Big]\Bigg)+\mathcal{O}(\betac^7)\Bigg]\\
    f_{\rm boost}^{(2)}(p) &= f_{\rm th}(p) P_2(\muc) \,\frac{\betac^2 p^2}{3\thetae^2}\,\Bigg[1+\frac{\betac^2}{14\thetae^2}\Big[p^2-7\gamma\thetae+14\thetae^2\Big]+\mathcal{O}(\betac^6)\Bigg] 
    \\
    f_{\rm boost}^{(3)}(p) &= f_{\rm th}(p) P_3(\muc) \,\frac{\betac^3 p^3}{15\thetae^3}\,\Bigg[1+\frac{\betac^2}{18\thetae^2}(p^2-9\gamma\thetae+27\thetae^2)
    +\mathcal{O}(\betac^7)\Bigg].
\end{split} \end{equation}
%--------------------------------------------------------------------------------
These expressions allows us to compare successive terms of the $\betac$ expansion with the `full' form of each angular multipole of the boosted distribution to determine the convergence of the SZ signal caused by the boosted Maxwell-J\"uttner distribution. This is shown graphically in Figure~\ref{fig:C_betac_exp_electron_boost}. Here we can immediately see that, even at high values of $\betac$ (displayed here at $\betac=0.1$) the signals very rapidly converge, requiring only a few terms before they are graphically indistinguishable from the `full' signal.

The largest divergence is seen in the monopole signal, wherein the leading order term contains no $\betac$ component and as such shows strong differences at large $\betac$. It is also worth noticing, in comparison to the CNSN method, that here for each multipole, even the first order term each multipole emerges at has broadly the same shape as the full distribution. That is, even at $\mathcal{O}(\betac^2)$ the quadrupole term of the kinematic boost method generates a shape with a null and two turning points in contrast to the CNSN quadrupole seen in e.g., Fig.~\ref{fig:5b_boost_vs CNSN}.

In conclusion, we can assert that the $\betac$ expansion does not cause the divergence between the CNSN model of the kinematic boost and the boosted Maxwell-J\"uttner distributiion method presented in this paper. Moreover, we can confirm that even at high $\betac$, the method presented here converges rapidly and thus low-order $\betac$ expansions could be used for the future calculation of these quantities.

%--------------------------------------------------------------------------------
%--------------------------------------------------------------------------------

%--------------------------------------------------------------------------------

%\bsp
\label{lastpage}
\end{document}